\documentclass[aps,eqsecnum,preprint,floats,epsf,epsfig,nofootinbib,letter]{revtex4}
\usepackage{epsfig}

\begin{document}
\def\bea{\begin{eqnarray}}
\def\eea{\end{eqnarray}}
\def\nc{N_c^{\rm eff}}
\def\vp{\varepsilon}
\def\drho{\bar\rho}
\def\deta{\bar\eta}
\def\a{{\cal A}}
\def\B{{\cal B}}
\def\c{{\cal C}}
\def\d{{\cal D}}
\def\e{{\cal E}}
\def\p{{\cal P}}
\def\t{{\cal T}}
\def\up{\uparrow}
\def\dw{\downarrow}
\def\vma{{_{V-A}}}
\def\vpa{{_{V+A}}}
\def\smp{{_{S-P}}}
\def\spp{{_{S+P}}}
\def\J{{J/\psi}}
\def\ov{\overline}
\def\Lqcd{{\Lambda_{\rm QCD}}}
\def\pr{{Phys. Rev.}~}
\def\prl{{Phys. Rev. Lett.}~}
\def\pl{{Phys. Lett.}~}
\def\np{{Nucl. Phys.}~}
\def\zp{{Z. Phys.}~}
\def\lsim{ {\ \lower-1.2pt\vbox{\hbox{\rlap{$<$}\lower5pt\vbox{\hbox{$\sim$}
}}}\ } }
\def\gsim{ {\ \lower-1.2pt\vbox{\hbox{\rlap{$>$}\lower5pt\vbox{\hbox{$\sim$}
}}}\ } }

\font\el=cmbx10 scaled \magstep2{\obeylines\hfill July, 2008}

\vskip 1.5 cm
\begin{center}{\large \bf Form Factors of $B_{u,d,s}$ Decays into
P-Wave Axial-Vector Mesons in the Light-Cone Sum Rule Approach}
\end{center}

\bigskip
\centerline{\bf Kwei-Chou Yang}
\centerline{Department of Physics, Chung Yuan Christian University}
\centerline{Chung-Li 320, Taiwan}
\bigskip
\bigskip
\bigskip
\bigskip

\small \centerline{\bf Abstract} We calculate the vector and axial-vector form
factors of $B_{u,d,s}$ decays into P-wave axial-vector mesons in the light-cone
sum rule approach. For the sum rule results, we have included corrections of
order $m_A/m_b$, where $m_A$ is the mass of the axial-vector meson $A$. The
results are relevant to the light-cone distribution amplitudes of the
axial-vector mesons. It is important to note that, owing to the $G$-parity, the
chiral-even two-parton light-cone distribution amplitudes of the $^3P_1$
($^1P_1$) mesons are symmetric (antisymmetric) under the exchange of quark and
anti-quark momentum fractions in the SU(3) limit. For chiral-odd light-cone
distribution amplitudes, it is the other way around. The predictions for decay
rates of $B_{u,d,s} \rightarrow A e \nu_e$ are also presented.
\bigskip
\small

\pagebreak

\section{Introduction}

The inclusive and exclusive $B$ decays provide potentially stringent test of
the Standard Model. Although the inclusive rare decays are theoretically clean,
they are a challenge for measurements at $B$ factories. The exclusive processes
may easily be accessible for experiments, but knowledge of form factors is
required.  The production of the axial-vector mesons has been seen in charmful
$B$ decays: $B\to J/\psi K_1(1270)$ and $B\to Da_1(1260)$ \cite{PDG}. As for
charmless hadronic $B$ decays, $B^0\to a_1^\pm(1260)\pi^\mp$ are the first
modes measured by BABAR and Belle \cite{BaBara1pi,Bellea1pi,BaBara1pitime}.
Information for weak phase $\alpha\equiv \arg(-V_{td}V_{tb}^*/V_{ud}V_{ub}^*)$
can be extracted from their time-dependent measurement or by relating these
decays with corresponding $\Delta S=1$ decays. BABAR has further reported the
observation of the decays $\ov B^0\to b_1^\pm\pi^\mp,b_1^+K^-,
K_1^-(1270)\pi^+,K_1^-(1400)\pi^+,a_1^+K^-$ and $B^-\to
b_1^0\pi^-,b_1^0K^-,a_1^0\pi^-,a_1^-\pi^0, a_1^-\ov
K^0,f_1(1285)K^-,f_1(1420)K^-$ \cite{Aubert:2007xd,Aubert:2007kpb}. Very
recently, $B^- \to K_1^-(1270) \phi, K_1^-(1400) \phi$ have been observed by
BABAR \cite{Aubert:2008zz}. Using the QCD factorization approach, we have
studied charmless two-body $B$ decays involving one or two axial-vector
meson(s) in the final state \cite{Yang:2007sb,Cheng:2007mx,Cheng:2008gx}.

In the quark model, two lowest nonets of $J^P=1^+$ axial-vector mesons are
expected as the orbitally excited $q\bar q^\prime$ states. In terms of the
spectroscopic notation $n^{2S+1}L_J$, where the radial excitation is denoted by
the principle number $n$, there are two types of the lowest $p$-wave mesons,
namely, $1^3P_1$ and $1^1P_1$. These two nonets have distinctive $C$ quantum
numbers, $C=+$ and $C=-$, respectively. Experimentally, the $J^{PC}=1^{++}$
nonet consists of $a_1(1260)$, $f_1(1285)$, $f_1(1420)$ and $K_{1A}$, while the
$1^{+-}$ nonet contains $b_1(1235)$, $h_1(1170)$, $h_1(1380)$ and $K_{1B}$. The
physical mass eigenstates $K_1(1270)$ and $K_1(1400)$ are mixtures of $K_{1A}$
and $K_{1B}$ states owing to the mass difference of the strange and non-strange
light quarks.

In QCD language a real hadron should be described in terms of a set of Fock
states for which each state has the same quantum number as the hadron. Because of G-parity, the decay constant for the {\it local} axial-vector ({\it local}
tensor) current coupling to the $^1 P_1$ ($^3 P_1$) state vanishes in the SU(3)
limit. However the constituent partons within a hadron are actually
non-localized. Projecting the axial-vector meson along the light-cone, due to
the G-parity the chiral-even light-cone distribution amplitudes (LCDAs) of a
$1^1P_1$ ($1^3P_1$) meson defined by the {\it nonlocal} axial-vector current is
antisymmetric (symmetric) under the exchange of $quark$ and $anti$-$quark$
momentum fractions in the SU(3) limit, whereas the chiral-odd LCDAs defined by
the {\it non-local} tensor current  are symmetric (antisymmetric)
\cite{Yang:2007zt,Yang:2005gk}. The large magnitude of the first Gegenbauer
moment of the mentioned antisymmetric LCDAs can have large impact on $B$ decays
involving a $1^3P_1$ or/and $1^1P_1$ meson(s). The related phenomenologies are
thus interesting \cite{Hatanaka:2008xj,Yang:2007sb,Cheng:2007mx}.

In this paper we present the first complete analysis for the form factors of
the $B_{u,d,s}$ decays into light axial-vector mesons ($A$) via the
vector/axial-vector current in the light-cone sum rule (LCSR) approach, where
$A$ is the light $P$-wave meson which can be the $1^3P_1, 1^1P_1$ or their
mixture state. The method of light-cone sum rules has been widely used in the
studies of nonperturbative processes, including weak baryon decays
\cite{Balitsky:1989ry}, heavy meson decays \cite{Chernyak:1990ag}, and heavy to
light transition form factors \cite{Belyaev:1993wp,Ball:1997rj,Ball:2004rg}.
Using the traditional QCD sum rule approach \cite{SVZ}, where the three-point
correlation function is considered, $B\to a_1$ form factors were calculated in
Ref.~\cite{Aliev:1999mx}. The BABAR measurement of $\overline B^0\to
a_1^+\pi^-$ \cite{BaBara1pitime} favors $V_0^{Ba_1}(0)\approx 0.30$
\cite{Yang:2007sb,Cheng:2007mx}, which is in good agreement with the LCSR
result that we obtain here.  The value given in Ref.~\cite{Aliev:1999mx} is a
little small but still consistent with the data within the errors. $B\to A$
form factors were studied in Ref. \cite{CCH} by using the light-front quark
model. Nevertheless, it is found to be $V^{Ba_1}_0(0)=0.13$ in the light-front
quark model calculation. It is interesting to note that very recently Wang
\cite{Wang:2008bw} used the $B$ meson light-cone sum rule approach to calculate
$B\to a_1$ form factors.

The rest of the paper is organized as follows. The definitions of decay
constants and $B_{q} \to A$ form factors are given in Sec.~\ref{sec:ff}. In
Sec.~\ref{sec:lcsr} we derive the light-cone sum rules for the relevant form
factors. The numerical results for form factors are given in
Sec.~\ref{sec:result}, where we also give the predictions for decay rates of
$B_{u,d,s}\to A e \nu_e$. A brief summary is given in Sec.~\ref{sec:summary}. The
relevant expressions for two-parton and three-parton LCDAs are collected in
Appendix \ref{app:2da-def}, an alternative definition for the form factors is
given in Appendix \ref{app:formfactor2}, and the formula for semileptonic
$B_{u,d,s}\to A e \nu_e$ decays is presented in Appendix \ref{app:decay-amplitude}.

\section{Definitions of decay constants and form factors}\label{sec:ff}

The G-parity\footnote{Here the idea of G-parity is extended to U-spin and
V-spin multiplets.} conserving decay constants of the axial-vector mesons are
defined as
\begin{eqnarray}
  \langle 1^3P_1(P,\lambda)|\bar q_1(0) \gamma_\mu \gamma_5 q_2(0)|0\rangle
  & = & if_{^3P_1}\, m_{^3P_1} \,  \epsilon^{*(\lambda)}_\mu
   \,, \label{eq:decay-const1.1}\\
 \langle 1^1P_1(P,\lambda)|
  \bar q_1(0) \sigma_{\mu\nu}\gamma_5 q_2(0)
   |0\rangle
 & = & f_{^1P_1}^{\perp}\,
(\epsilon^{*(\lambda)}_{\mu} P_{\nu} - \epsilon_{\nu}^{*(\lambda)}
P_{\mu}),\label{eq:decay-const1.2}
\end{eqnarray}
where $f_{^3P_1}$ is scale-independent, but $f_{^1P_1}^{\perp}$ is
scale-dependent \cite{Yang:2007zt,Yang:2005gk}. On the other hand, we define
the G-parity violating decay constants to be
\begin{eqnarray}\label{eq:decay-const2}
 \langle 1^3P_1(P,\lambda)|
  \bar q_1(0) \sigma_{\mu\nu}\gamma_5 q_2(0)
   |0\rangle
 & = & f_{^3P_1} a_0^{\perp,^3P_1} \,
(\epsilon^{*(\lambda)}_{\mu} P_{\nu} - \epsilon_{\nu}^{*(\lambda)}
P_{\mu}), \\
 \langle 1^1P_1(P,\lambda)|\bar q_1(0) \gamma_\mu \gamma_5 q_2(0)|0\rangle
 &  = & if_{^1P_1}^\perp(1~{\rm GeV}) a_0^{\parallel,^1P_1} \, m_{^1P_1} \,
 \epsilon^{*(\lambda)}_\mu
   \,,
 \end{eqnarray}
where $a_0^{\perp,^{3}P_1}$ and $a_0^{\parallel,^{1}P_1}$, which are
respectively the zeroth Gegenbauer moments of $\Phi_\perp^{^{3}P_1}$ and
$\Phi_\parallel^{^{1}P_1}$, are zero in the SU(3) limit. The definitions for
the LCDAs $\Phi_\perp^{^{3}P_1}$ and $\Phi_\parallel^{^{1}P_1}$ are collected
in Appendix~\ref{app:2da-def}. We further define $f_{^3\! P_1}^\perp=f_{^3\!
P_1}$ and $f_{^1\! P_1}=f_{^1\! P_1}^\perp(\mu=1~{\rm GeV})$ in the present
study as did in Ref.~\cite{Yang:2007zt}. In the present work,  the $G$-parity
violating parameters, e.g. $a_1^{\parallel,K_{1A}}, a_{0,2}^{\perp,K_{1A}},
a_1^{\perp,K_{1B}}$ and $a_{0,2}^{\parallel,K_{1B}}$, are considered for mesons
containing a strange quark. Note that for $G$-parity violating quantities,
their signs have to be flipped from mesons to anti-mesons.

The semileptonic form factors for the $\bar B_{q}\to A$ transition are
defined as
\begin{eqnarray} \label{eq:FF}
\langle{A}(P, \lambda)|A_\mu|{\overline B_{q}} (p_B)\rangle
 &=& i \frac{2}{m_{B_q} + m_{A}} \epsilon_{\mu\nu\alpha\beta} \epsilon_{(\lambda)}^{*\nu}
 p_B^\alpha P^{\beta} A^{B_q A}(q^2),
\nonumber \\
 \langle A (P,\lambda)|V_\mu|{\overline B_q}(p_B)\rangle
 &=& - \Bigg\{ (m_{B_q} + m_{A}) \epsilon^{(\lambda)*}_{\mu} V_1^{B_q A}(q^2)
 - (\epsilon^{(\lambda)*} \cdot p_B)
(p_B + P)_\mu \frac{V_2^{B_q A}(q^2)}{m_{B_q} + m_{A}}
\nonumber \\
&& - 2 m_{A} \frac{\epsilon^{(\lambda)*}\cdot p_B}{q^2} q^\mu
\left[V_3^{B_q A}(q^2) - V_0^{B_q A}(q^2)\right]\Bigg\}, \nonumber \\
 \end{eqnarray}
where $q = p_B - P$, $V_3^{B_q A}(0) = V_0^{B_q A}(0)$,
 \begin{eqnarray}
 V_3^{B_q A}(q^2)&=& \frac{m_{B_q} + m_{A}}{2 m_{A}} V_1^{B_q A}(q^2) - \frac{m_{B_q} -
 m_{A}}{2 m_{A}} V_2^{B_q A}(q^2)\,,
 \label{eq:V30}
 \end{eqnarray}
and we adopt the convention $\epsilon^{0123}=-1$.  An alternative definition
for the form factors is given in Appendix \ref{app:formfactor2}.

\section{The light-cone sum rules}\label{sec:lcsr}

We consider the following two-point correlation function, which is sandwiched
between the vacuum and transversely polarized $A$ meson (in this section
$A\equiv$ a pure $1^3P_1$ or $1^1P_1$ state), to calculate the form factors:

\begin{eqnarray}
&& i\int d^4x e^{iq x} \langle A(P,\perp)|T [\bar
q_1(x)\gamma_\mu(1-\gamma_5) b(x)\, j_{B_{q_2}}^\dagger(0)]|0\rangle \nonumber\\
&& ~~ = -{\bf V_1}(q^2) \epsilon^{*(\perp)}_\mu
 + {\bf V_2}(q^2) (\epsilon^{*(\perp)} q) (2P+q)_\mu
 + {\bf V}(q^2) \frac{\epsilon^{*(\perp)} q}{q^2} q_\mu  \nonumber\\
&& ~~~~~\, -i {\bf A}(q^2)\epsilon_{\mu\nu\rho\sigma} \epsilon^{*\nu}_{(\perp)} q^\rho P^\sigma \,,
\label{eq:correlator-1}
\end{eqnarray}
where $p_B^2=(P+q)^2$, $P$ is the momentum of the $A$ meson, and $j_B=i \bar
q_2 \gamma_5 b$ (with $q_{1(2)} \equiv u, d$ {\rm or} $s$) is the
interpolating current for the ${B_{q_2}}$ meson, so that
\begin{equation}\label{eq:B-decay-cont}
\langle 0|j_{B_{q_2}}(0)|\bar B_{q_2}(p_B)\rangle = \frac{f_{B_{q_2}} m_{B_{q_2}}^2}{m_b+m_{q_2}}.
\end{equation}

In the region of sufficiently large virtualities: $m_b^2 - p_B^2\gg
\Lambda_{\rm QCD}m_b$ with $q^2$ being small and positive, the operator product
expansion is applicable in Eq.~(\ref{eq:correlator-1}), so that for an
energetic $A$ meson the correlation function in Eq.~(\ref{eq:correlator-1}) can
be represented in terms of the LCDAs of the $A$ meson:
\begin{eqnarray}
\lefteqn{i\int d^4x e^{iq x} \langle A(P,\perp)|T [\bar
q_1(x)\gamma_\mu(1-\gamma_5) b(x)\, j_B^\dagger(0)]|0\rangle }
\hspace*{1.0cm} \nonumber\\
  & & = \int_0^1 du \frac{-i}{(q+ k)^2-m_b^2}
 {\rm Tr} \Big[\gamma_\mu(1-\gamma_5) (\not\! q + \not\! k +m_b)\gamma_5
M_\perp^{A} \Big] \Bigg|_{k=uE n_-}
   \cr
 && ~~~ +{\cal O}\bigg(\frac{m_{A}^2}{E^2}\bigg) \,,~~~\label{eq:green-fn-2}
\end{eqnarray}
where $E=|\vec{P}|$, $P^\mu =En_-^\mu + m_{A}^2 n_+^\mu/(4E) \simeq E n_-^\mu$
with two light-like vectors $n_-^\mu=(1,0,0,-1)$ and $n_+^\mu=(1,0,0,1)$
satisfying $n_- n_+=2$ and $n_-^2=n_+^2=0$. Here, $E\sim m_b$, and we have
assigned the momentum of the $q_1$-quark in the $A$ meson to be
 \begin{eqnarray}
 k^\mu = u E n_-^\mu +k_\perp^\mu + \frac{k_\perp^2}{4 uE}n_+^\mu\,,
\end{eqnarray}
where $k_\perp$ is of order $\Lambda_{\rm QCD}$. Here $u$ is the momentum
fraction carried by the quark  in the axial-vector meson. In
Eq.~({\ref{eq:green-fn-2}), to calculate contributions in the momentum space,
we have used the following substitution
\begin{equation}\label{eq:x-to-k}
x^\mu \to -i \frac{\partial}{\partial k_{\mu}}\simeq -i \Bigg(
\frac{n_+^\mu}{2E}\frac{\partial}{\partial u} +
\frac{\partial}{\partial k_{\perp\, \mu}}\Bigg)\,,
\end{equation}
to the Fourier transform for
 \begin{eqnarray}\label{eq:fourier}
  &&  \langle A(P,\lambda)|\bar q_{1\,\alpha}(x) \, q_{2\, \delta}(0)|0\rangle
= -\frac{i}{4} \, \int_0^1 du \,  e^{i u \, P x}
\nonumber\\[0.1cm]
  && \quad \times\,\Bigg\{ f_A m_A \Bigg[
  \not\! P\gamma_5 \, \frac{\epsilon^*_{(\lambda)} x}{Px} \,
  \Bigg(\Phi_\parallel(u) + \frac{m_A^2 x^2}{16} \mathbf{A}_\parallel^2\Bigg)
  +\Bigg( \not\! \epsilon^* -\not\! P \frac{\epsilon^*_{(\lambda)} x}{Px}\Bigg)\,
   \gamma_5 g_\perp^{(a)}(u) \nonumber\\
  && \qquad - \not\! x\gamma_5 \frac{\epsilon^*_{(\lambda)} x}{2(Px)^2}
  m_A^2 \bar g_3(u) +
 \epsilon_{\mu\nu\rho\sigma} \,
    \epsilon^*_{(\lambda)}{}^\mu  p^{\rho} x^\sigma \, \gamma^\mu
    \, \frac{g_\perp^{(v)}(u)}{4}\Bigg]
\nonumber \\[0.1em]
  && \qquad
  + \,f^{\perp}_A \Bigg[
  \frac{1}{2}\bigg( \! \not\! P\not\!\epsilon^*_{(\lambda)}-
  \not\!\epsilon^*_{(\lambda)} \not\! P  \bigg) \gamma_5
 \Bigg(\Phi_\perp(u) + \frac{m_A^2 x^2}{16} \mathbf{A_\perp}^2\Bigg)
 \nonumber\\
 && \qquad
 \,\,\, -
   \frac{1}{2}\bigg( \! \not\! P\not\! x- \not\! x \not\! P  \bigg)
   \gamma_5 \frac{\epsilon^*_{(\lambda)} x}{(Px)^2} m_A^2 \bar
   h_\parallel^{(t)} (u)
 -\frac{1}{4}\bigg( \! \not\!\epsilon^*_{(\lambda)} \not x-
  \not x \not\!\epsilon^*_{(\lambda)} \bigg) \gamma_5
  \frac{m_A^2 }{Px} \bar h_3 (u)\nonumber\\
 && \qquad\,\,\, +i \Big(\epsilon^*_{(\lambda)} x\Big) m_A^2
 \gamma_5
 \frac{h^{(p)}_\parallel (u)}{2}
  \Bigg]\Bigg\}_{\delta\alpha}\,,
 \end{eqnarray}
where $x^2 \neq 0$,
\begin{eqnarray}
 \bar g_3 (u) &=& g_3(u) +\Phi_\parallel -2 g_\perp^{(a)}(u),\nonumber\\
 \bar h_\parallel^{(t)} &=& h_\parallel^{(t)}
   - \frac{1}{2} \Phi_\perp(u) -\frac{1}{2} h_3(u) , \nonumber\\
 \bar h_3(u) &=& h_3(u) -\Phi_\perp(u).
\end{eqnarray}
The detailed definitions for the relevant two-parton LCDAs are collected in
Appendix~\ref{app:2da-def}. In Eq. (\ref{eq:x-to-k}) the term of order
$k_\perp^2$ is omitted in the calculation. Consequently, we can obtain the
light-cone projection operator of the $A$ meson in the momentum space,
\begin{equation}
  M_{\delta\alpha}^A =  M_{\delta\alpha}{}_\parallel^A +
   M_{\delta\alpha}{}_\perp^A \,,
\label{app:rhomeson2}
\end{equation}
where $M^A_{\delta\alpha}{}_\parallel$ and $M^A_{\delta\alpha}{}_\perp$ are
the longitudinal and transverse projectors, respectively. The longitudinal
projector which projects the longitudinal component of the axial-vector meson
is given by
 \begin{eqnarray} \label{eq:AproL}
M^A_\parallel &=& -i\frac{f_A}{4} \, \frac{m_A(\epsilon^*_{(\lambda)} n_+)}{2} \Bigg\{
 \not\! n_- \gamma_5\,\Phi_\parallel(u)
 - \frac{f_A^\perp}{f_A} \frac{m_A}{E}
 \, \Bigg[-\frac{i}{2}\,\sigma_{\mu\nu}\gamma_5 \,  n_-^\mu  n_+^\nu \,
 h_\parallel^{(t)}(u)
\nonumber\\
&& \hspace*{-0.0cm} - \,i E\int_0^u dv \,(\Phi_\perp(v) -
h_\parallel^{(t)}(v)) \
     \sigma_{\mu\nu} \gamma_5 n_-^\mu
     \, \frac{\partial}{\partial k_\perp{}_\nu}
  +\gamma_5\frac{h_\parallel'^{(p)}(u)}{2}\Bigg]\, \Bigg|_{k=u p}
  +{\cal O}\Bigg(\frac{m_A^2}{E^2}\Bigg) \Bigg\}\,, \ \ \
\end{eqnarray}
 and the transverse projector reads
 \begin{eqnarray} \label{eq:AproT}
M^A_\perp &=& i\frac{f^{\perp}_A}{4} E \Bigg\{\not\!
\epsilon^{*(\lambda)}_\perp\not\! n_- \gamma_5 \,
   \Phi_\perp(u)\nonumber\\
&&  - \frac{f_A}{f_A^\perp}\frac{m_A}{E} \,\Bigg[ \not\!
\epsilon^{*(\lambda)}_\perp\gamma_5 \, g_\perp^{(a)}(u) -  \, E\int_0^u dv\,
(\Phi_\parallel(v) - g_\perp^{(a)}(v)) \not\! n_-\gamma_5 \,
\epsilon^{*(\lambda)}_{\perp\mu} \,\frac{\partial}{\partial
         k_{\perp\mu}}
\cr && + \,i \varepsilon_{\mu\nu\rho\sigma} \,
       \gamma^\mu \epsilon_\perp^{*(\lambda)\nu} \,  n_-^\rho
         \left( n_+^\sigma \,{g_\perp^{(v)\prime}(u)\over 8}-
          E\,\frac{g_\perp^{(v)}(u)}{4} \, \frac{\partial}{\partial
         k_\perp{}_\sigma}\right)
 \Bigg]
 \, \Bigg|_{k=up} + {\cal O}\Bigg(\frac{m_A^2}{E^2}\Bigg) \Bigg\}\,,
\end{eqnarray}
where the {\it exactly} longitudinal and transverse polarization vectors of
the axial-vector meson, independent of the coordinate variable $x$, are
defined as
\begin{equation}\label{eq:exact-polarization}
\epsilon^{*(0)\mu}= \frac{E}{m_{A}} \Bigg[ \Bigg(1-\frac{m_{A}^2}{4E^2}\Bigg)
n_-^\mu
 - \frac{m_{A}^2}{4 E^2} n_+^\mu\Bigg], \quad
\epsilon^{*(\lambda)\mu}_\perp \equiv
\Bigg(\epsilon^{*(\lambda)\mu} - \frac{\epsilon^{*(\lambda)}
n_+}{2}\,n_-^\mu- \frac{\epsilon^{*(\lambda)} n_-}{2}\,n_+^\mu \Bigg)
\delta_{\lambda,\pm1}\,.
\end{equation}
Here we have assumed that the meson moves along the $n_-^\mu$ direction.  A
similar discussion about the projection operator for the vector meson can be
found in Ref.~\cite{Beneke:2000wa}. From the expansion of the transverse
projection operator, one can find that contributions arising from
$g_\perp^{(a)}, \Phi_\parallel -g_\perp^{(a)}, g_\perp^{(v)\prime}$, and
$g_\perp^{(v)}$ are suppressed by $m_{A}/E$ as compared with the term with
$\Phi_\perp$. Note that in Eq.~(\ref{eq:green-fn-2}) the derivative with
respect to the transverse momentum acts on the hard scattering amplitude
before the collinear approximation is taken. Note also that applying equations
of motion, the twist-three two-parton LCDAs can be expressed in terms of
leading-twist and twist-three three-parton light-cone distribution amplitudes.
One can thus understand from Eqs.~(\ref{eq:AproL}) and (\ref{eq:AproT}) that
the expansion parameter in the light-cone sum rules should be $m_A/m_b$,
instead of twist.

At the quark-gluon level, after performing the integration of
Eq.~(\ref{eq:green-fn-2}), the results up to ${\cal O}(m_A/m_b)$ read
\begin{eqnarray}
 \bf{V_1}^{\rm QCD}&=& - \frac{m_b^2 f_{A}^\perp}{2} \int_0^1 \frac{du}{u}
\Bigg\{ \frac{1} {m_b^2 -up_B^2 -\bar u q^2}
  \Bigg[\frac{m_b^2 -q^2}{m_b^2} \Phi^\perp(u)
 -\Bigg( \frac{ m_{A}f_{A} }{m_b f_{A}^\perp } \Bigg)
  2 ug^{(a)}_\perp(u)
 \Bigg] \Bigg\} , ~~~~~~~
\end{eqnarray}
\begin{eqnarray}
  \bf{V_2}^{\rm QCD}&=& - \frac{m_b^2 f_{A}^\perp}{2} \int_0^1 du \Bigg\{
  \frac{1}
  {m_b^2 -up_B^2 -\bar u q^2} \frac{\Phi^\perp(u)}{m_b^2} -
  \frac{ m_{A}f_{A} }{ m_b f_{A}^\perp }
  \frac{2 \Phi_a(u)}{(m_b^2-up_B^2 -\bar u q^2)^2}
   \Bigg\} ,~~~~~
\end{eqnarray}
\begin{eqnarray}
  \bf{V}^{\rm QCD}&=&  \frac{q^2 m_b^2 f_{A}^\perp}{2} \int_0^1 du
  \Bigg\{\frac{1}
  {m_b^2 -up_B^2 -\bar u q^2} \frac{\Phi^\perp(u)}{m_b^2} -
  \frac{ m_{A}f_{A} }{ m_b f_{A}^\perp }
  \frac{2 \Phi_a(u)}{(m_b^2-up_B^2 -\bar u q^2)^2}
   \Bigg\} , ~~~~~~~
\end{eqnarray}
\begin{eqnarray}
  \bf{A}^{\rm QCD}&=& -m_b^2 f_{A}^\perp \int_0^1 du \Bigg\{
  \frac{1}
  {m_b^2 -up_B^2 -\bar u q^2} \frac{\Phi^\perp(u)}{m_b^2} -
  \frac{ m_{A}f_{A} }{ m_b f_{A}^\perp }
  \frac{g_\perp^{(v)}(u)/2}{(m_b^2-up_B^2 -\bar u q^2)^2}
   \Bigg\} ,~~~~~
\end{eqnarray}
where $\Phi_a(u)\equiv \int_0^u dv\, (\Phi_\parallel(v) - g_\perp^{(a)}(v))$
and $\bar u \equiv 1-u$. Note that here the contributions due to the {\it
explicit} three-parton LCDAs are suppressed by ${\cal O}(m_A^2/m_b^2)$ as
compared with the term involving $\Phi_\perp$.

We have given the results for $\mathbf{V_1}, \mathbf{V_2}$ and $\mathbf{A}$
from the hadron and quark-gluon points of view. Thus, for instance, for the
form factor $V_1$ the contribution due to the lowest-lying $A$ meson can be
further approximated with the help of quark-hadron duality:
\begin{equation}
  V_1(q^2)\cdot \frac{m_{B_{q_2}}+m_A}{m^2_{B_{q_2}}-p_B^2} \cdot
        \frac{m_{B_{q_2}}^2 f_{B_{q_2}}}{m_b + m_{q_2}} = \frac{1}{\pi}\int_{m_b^2}^{s_0}
        \frac{{\rm Im}\mathbf{V_1}^{\rm QCD}(s,q^2)}{s-p_B^2}ds\,,
\end{equation}
where $s_0$ is the excited state threshold. After applying the Borel transform
$p_B^2 \to M^2$  \cite{SVZ,Chernyak:1990ag,Ball:1997rj} to the above equation,
we obtain
\begin{eqnarray}
 V_1(q^2) = \frac{(m_b + m_{q_2})}{(m_{B_{q_2}}+m_A) m_{B_{q_2}}^2 f_{B_{q_2}}}
 e^{m_{B_{q_2}}^2/M^2} \frac{1}{\pi}\int_{m_b^2}^{s_0}
       e^{s/M^2} {\rm Im}\mathbf{V_1}^{\rm QCD}(s,q^2) ds\,.
 \end{eqnarray}

We obtain the light-cone sum rule results:
\begin{eqnarray}
 V^{{B_{q_2}}A}_1(q^2)
 &=& - \frac{(m_b + m_{q_2})m_b^2 f_{A}^\perp }{(m_{B_{q_2}}+m_A)m_{B_{q_2}}^2
 f_{B_{q_2}}} e^{(m_{B_{q_2}}^2 -m_b^2)/M^2}
 \int_0^1 du \Bigg\{ \frac{1}{u} e^{\bar u(q^2 -m_b^2)/(uM^2)} \theta[c(u,s_0)]\nonumber\\
& & \times \Bigg(\Phi^\perp(u) \frac{m_b^2-q^2}{2u m_b^2}
 -\frac{m_{B_{q_2}}+m_A}{m_{B_{q_2}}}\frac{ m_{A}f_{A} }{m_b f_{A}^\perp }
 g^{(a)}_\perp(u) \Bigg) \Bigg\} ,
\end{eqnarray}
\begin{eqnarray}
 V^{{B_{q_2}}A}_2(q^2)
&=& - \frac{(m_b + m_{q_2}) (m_{B_{q_2}}+m_A) f_{A}^\perp }{2 m_{B_{q_2}}^2 f_{B_{q_2}}}
 e^{(m_{B_{q_2}}^2 -m_b^2)/M^2}
\int_0^1 du \Bigg\{ \frac{1}{u} e^{\bar u(q^2 -m_b^2)/(uM^2)}
        \Bigg[ \Phi^\perp(u) \theta[c(u,s_0)]\nonumber\\
& & \ \ -  \frac{m_{B_{q_2}}}{m_{B_{q_2}}+m_A}\frac{2m_{A}m_b f_{A}}{uM^2 f_{A}^\perp } \Phi_a(u)
 \Bigg(\theta[c(u,s_0)] + uM^2 \delta[c(u,s_0)] \Bigg) \Bigg] \Bigg\} ,
\end{eqnarray}
\begin{eqnarray}
 V_0^{{B_{q_2}}A}(q^2)
&=& V^{{B_{q_2}}A}_3(q^2) - \frac{q^2 (m_b + m_{q_2}) f_{A}^\perp }
{4 m_{B_{q_2}}^2 m_A f_{B_{q_2}}}   e^{(m_{B_{q_2}}^2
-m_b^2)/M^2} \int_0^1 du \Bigg\{ \frac{1}{u} e^{\bar u(q^2 -m_b^2)/(uM^2)}
        \Bigg[ \Phi^\perp(u) \theta[c(u,s_0)]\nonumber\\
& & \ \ -  \frac{2m_{A}m_b f_{A}}{uM^2 f_{A}^\perp } \Phi_a(u)
 \Bigg(\theta[c(u,s_0)] + uM^2 \delta[c(u,s_0)] \Bigg) \Bigg] \Bigg\} ,
\end{eqnarray}
\begin{eqnarray}
 A^{{B_{q_2}}A}(q^2)
 &=& - \frac{(m_b + m_{q_2}) (m_{B_{q_2}}+m_A) f_{A}^\perp }{2m_{B_{q_2}}^2 f_{B_{q_2}}}
  e^{(m_{B_{q_2}}^2 -m_b^2)/M^2}
\int_0^1 du \Bigg\{ \frac{1}{u} e^{\bar u(q^2 -m_b^2)/(uM^2)}
        \Bigg[ \Phi^\perp(u) \theta[c(u,s_0)]\nonumber\\
& & \ \ -  \frac{m_{B_{q_2}}}{m_{B_{q_2}}+m_A}\frac{m_{A} m_b f_{A}}{2
 uM^2 f_{A}^\perp } g^{(v)}_\perp(u)
 \Bigg(\theta[c(u,s_0)] + uM^2 \delta[c(u,s_0)] \Bigg) \Bigg] \Bigg\} ,
\end{eqnarray}
where $V^{{B_{q_2}}A}_3(q^2)$ is given by Eq.~(\ref{eq:V30}), and
 $c(u,s_0)=us_0 -m_b^2 + (1-u) q^2$.

\section{Results}\label{sec:result}

\subsection{Input parameters}\label{subsec:inputs}

In this subsection we shall briefly summarize the relevant input parameters, which are collected in
Tables~\ref{tab:Mass-DecayConst}, \ref{tab:Gegenbauer} and  \ref{tab:inputs}.
 The masses of $u$ and $d$ quarks are neglected.


The physical states $K_1(1270)$ and $K_1(1400)$ are the mixtures of the
$K_{1A}$ and $K_{1B}$. $K_{1A}$ and $K_{1B}$ are not mass eigenstates, and
they can be mixed together due to the strange and nonstrange light quark mass
difference. Their relations can be written as
 \begin{eqnarray}
 \label{eq:mixing}
 |\bar K_1(1270)\rangle &=& |\bar K_{1A}\rangle\sin\theta_{K_1}+
  |\bar K_{1B}\rangle\cos\theta_{K_1}, \nonumber \\
 |\bar K_1(1400)\rangle &=& |\bar K_{1A}\rangle\cos\theta_{K_1} -
 |\bar K_{1B}\rangle\sin\theta_{K_1}.
 \end{eqnarray}
The sign ambiguity for $\theta_{K_1}$ is due to the fact that one can add
arbitrary phases to $|\bar K_{1A}\rangle$ and $|\bar K_{1B}\rangle$. This sign
ambiguity can be removed by fixing the signs for $f_{K_{1A}}$ and
$f_{K_{1B}}^\perp$, which do not vanish in the SU(3) limit. Following
Ref.~\cite{Yang:2007zt}, we adopt the convention: $f_{K_{1A}}>0$,
$f_{K_{1B}}^\perp>0$. Combining the analyses for the data of the decays $B\to
K_1 \gamma$ and $\tau^-\to K_1^-(1270)\nu_\tau$ \cite{Abe:2004kr,PDG}, the
mixing angle was found to be $\theta_{K_1}=-(34 \pm 13)^\circ$
\cite{Hatanaka:2008xj}.

Analogous to the $\eta-\eta'$ mixing in the pseudoscalar sector, the $1^3P_1$
states, $f_1(1285)$ and $f_1(1420)$, have mixing via
\begin{eqnarray}
 |f_1(1285)\rangle = |f_1\rangle\cos\theta_{^3P_1}+|f_8\rangle\sin\theta_{^3P_1},
 \quad |f_1(1420)\rangle =
 -|f_1\rangle\sin\theta_{^3P_1} +|f_8\rangle\cos\theta_{^3P_1} \,,
 \end{eqnarray}
and likewise the $1^1P_1$ states, $h_1(1170)$ and $h_1(1380)$, can be
mixed in terms of the pure octet $h_8$ and singlet $h_1$,
 \begin{eqnarray}
 |h_1(1170)\rangle = |h_1\rangle\cos\theta_{^1P_1}+|h_8\rangle\sin\theta_{^1P_1},
 \quad
 |h_1(1380)\rangle = -|h_1\rangle\sin\theta_{^1P_1} +|h_8\rangle\cos\theta_{^1P_1} \,.
 \end{eqnarray}
Using the Gell-Mann-Okubo mass formula \cite{Yang:2007zt}, we obtain the
mixing angles $\theta_{^1P_1}$ and $\theta_{^3P_1}$ to be
\begin{eqnarray} \label{eq:mixingangle}
 \theta_{^3P_1}=(23.6^{+17.0}_{-23.6})^\circ, \quad &&
\theta_{^1P_1}=(28.1^{+~9.8}_{-17.2})^\circ, \quad {\rm for}~\theta_{K_1}=-(34
\pm 13)^\circ.
 \end{eqnarray}
For $^3P_1$ states the decay constants $f_{f_1(1285)}^q$ and $f_{f_1(1420)}^q$
are defined by
\begin{eqnarray}
\langle 0| \bar q\gamma_\mu \gamma_5 q | f_1(1285)(P,\lambda)\rangle
 &=& -i
m_{f_1(1285)} f_{f_1(1285)}^q
\epsilon_\mu^{(\lambda)}\,,\label{eq:decay-def1}\\ \nonumber \langle 0| \bar
q\gamma_\mu \gamma_5 q | f_1(1420)(P,\lambda)\rangle
 &=& -i m_{f_1(1420)} f_{f_1(1420)}^q \epsilon_\mu^{(\lambda)}\,, \label{eq:decay-def2}
\end{eqnarray}
and for $1^1P_1$ states the tensor decay constants are
\begin{eqnarray}
\langle 0| \bar q\sigma_{\mu\nu} q | h_1(1170)(P,\lambda)\rangle
 &=& i f_{h_1(1170)}^{\perp,q}\,\epsilon_{\mu\nu\alpha\beta}
  \epsilon_{(\lambda)}^\alpha P^\beta\,, \label{eq:decay-def3}\\ \nonumber
\langle 0| \bar q\sigma_{\mu\nu} q | h_1(1380)(P,\lambda)\rangle
 &=& i f_{h_1(1380)}^{\perp,q}\,\epsilon_{\mu\nu\alpha\beta}
  \epsilon_{(\lambda)}^\alpha P^\beta\,. \label{eq:decay-def4}
\end{eqnarray}
The reader is referred to \cite{Yang:2007zt,Cheng:2007mx} for details.

%
\begin{table}[h]
\caption[]{Masses and decay constants for $1^3P_1$ and $1^1P_1$ states
obtained in the QCD sum rule calculation \cite{Yang:2007zt}.}
\label{tab:Mass-DecayConst}
\renewcommand{\arraystretch}{1.5}
\addtolength{\arraycolsep}{4pt}
$$
\begin{array}{|c|cc|}\hline\hline
{\rm State} & {\rm Mass}~ [{\rm GeV}]& {\rm Decay\ constant}~f_{^{3(1)}P_1}
[{\rm MeV}]
 \\ \hline
\begin{array}{c} a_1(1260)   \\ f_1 (1^3P_1) \\ f_8 (1^3P_1) \\ K_{1A}\end{array}&
\begin{array}{c} 1.23\pm 0.06\\ 1.28\pm 0.06 \\ 1.29\pm 0.05 \\ 1.31\pm 0.06\end{array}&
\begin{array}{c} 238\pm 10   \\ 245\pm 13    \\ 239\pm 13    \\ 250\pm 13\end{array}
 \\ \hline
\begin{array}{c} b_1(1235) \\ h_1 (1^1P_1) \\ h_8 (1^1P_1) \\ K_{1B}\end{array}&
\begin{array}{c} 1.21\pm 0.07\\ 1.23\pm 0.07\\ 1.37\pm 0.07\\ 1.34\pm 0.08\end{array}&
\begin{array}{c} 180\pm 8\\ 180\pm 12\\ 190\pm 10\\ 190\pm 10\end{array}
\\ \hline\hline
\end{array}
$$
\end{table}
%
\begin{table}[h]
\caption[]{Gegenbauer moments of $\Phi_\perp$ and $\Phi_\parallel$ for
$1^3P_1$ and $1^1P_1$ mesons, respectively, where $a_0^{\perp,K_{1A}}$ and
$a_0^{\parallel,K_{1B}}$ are updated from the $B\to K_1 \gamma$ analysis, and
$a_1^{\parallel,K_{1A}}, a_2^{\perp,K_{1A}}, a_2^{\parallel,K_{1B}}$, and
$a_1^{\perp,K_{1B}}$ are then obtained from Eq.~(141) in
Ref.~\cite{Yang:2007zt}.} \label{tab:Gegenbauer}
\renewcommand{\arraystretch}{1.8}
\addtolength{\arraycolsep}{0.4pt} {\small
$$
\begin{array}{|c|c|c|c|c|c|c|}\hline
 \mu & a_2^{\parallel, a_1(1260)}& a_2^{\parallel,f_1^{^3P_1}}
 & a_2^{\parallel,f_8^{^3P_1}} & a_2^{\parallel, K_{1A}}
 & \multicolumn{2}{|c|}{a_1^{\parallel, K_{1A}}}
 \\ \hline
\begin{array}{c} {\rm 1~GeV}   \\ {\rm 2.2~GeV} \end{array}&
\begin{array}{c} -0.02\pm 0.02 \\ -0.01\pm 0.01  \end{array}&
\begin{array}{c}  -0.04\pm 0.03 \\ -0.03 \pm 0.02 \end{array}&
\begin{array}{c} -0.07\pm 0.04\\ -0.05 \pm  0.03 \end{array}&
\begin{array}{c} -0.05\pm 0.03\\  -0.04 \pm 0.02 \end{array}&
\multicolumn{2}{c|}{\begin{array}{c} {-0.30^{+0.00}_{-0.20}}\\
{-0.25^{+0.00}_{-0.17}}\end{array}}
\\ \hline\hline
 \mu & a_1^{\perp, a_1(1260)}& a_1^{\perp,f_1^{^3P_1}}
 & a_1^{\perp,f_8^{^3P_1}} & a_1^{\perp, K_{1A}}
 & a_0^{\perp, K_{1A}}    & a_2^{\perp, K_{1A}}
 \\ \hline
\begin{array}{c} {\rm 1~GeV}   \\ {\rm 2.2~GeV} \end{array}&
\begin{array}{c} -1.04\pm 0.34 \\ -0.85\pm 0.28  \end{array}&
\begin{array}{c} -1.06\pm 0.36 \\ -0.86\pm 0.29   \end{array}&
\begin{array}{c} -1.11\pm 0.31 \\ -0.90\pm 0.25   \end{array}&
\begin{array}{c} -1.08\pm 0.48 \\ -0.88\pm 0.39   \end{array}&
\begin{array}{c}  0.27^{+0.03}_{-0.17}\\ 0.25^{+0.03}_{-0.16}
 \end{array}&
\begin{array}{c}  0.02\pm 0.21\\   0.01\pm 0.15   \end{array}
\\ \hline\hline
\mu & a_1^{\parallel, b_1(1235)}& a_1^{\parallel,h_1^{^1P_1}}
 & a_1^{\parallel,h_8^{^1P_1}} & a_1^{\parallel, K_{1B}}
 & a_0^{\parallel, K_{1B}}    & a_2^{\parallel, K_{1B}}
 \\ \hline
\begin{array}{c} {\rm 1~GeV}   \\   {\rm 2.2~GeV} \end{array}&
\begin{array}{c} -1.95\pm 0.35 \\  -1.61\pm 0.29  \end{array}&
\begin{array}{c} -2.00\pm 0.35 \\  -1.65\pm 0.29  \end{array}&
\begin{array}{c} -1.95\pm 0.35 \\  -1.61\pm 0.29  \end{array}&
\begin{array}{c} -1.95\pm 0.45 \\  -1.57\pm 0.37  \end{array}&
\begin{array}{c}  -0.19\pm 0.07 \\   -0.19\pm 0.07  \end{array}&
\begin{array}{c}  0.10^{+0.15}_{-0.19} \\    0.07^{+0.11}_{-0.14}
 \end{array}
\\ \hline\hline
 \mu & a_2^{\perp, b_1(1235)}& a_2^{\perp,h_1^{^1P_1}}
 & a_2^{\perp,h_8^{^1P_1}} & a_2^{\perp, K_{1B}}
 & \multicolumn{2}{|c|}{a_1^{\perp, K_{1B}}}
 \\ \hline
\begin{array}{c}  {\rm 1~GeV}  \\   {\rm 2.2~GeV} \end{array}&
\begin{array}{c}  0.03\pm 0.19 \\   0.02\pm 0.15  \end{array}&
\begin{array}{c}  0.18\pm 0.22 \\  0.14 \pm 0.17  \end{array}&
\begin{array}{c}   0.14\pm 0.22\\  0.11 \pm  0.17 \end{array}&
\begin{array}{c}  -0.02\pm 0.22\\ -0.02 \pm 0.17  \end{array}&
\multicolumn{2}{c|}{\begin{array}{c} 0.30^{+0.00}_{-0.33}\\
                                     0.24^{+0.00}_{-0.27}\end{array}}
\\ \hline
\end{array}
$$}
\end{table}
%
\begin{table}[ht!]
\centerline{\parbox{16cm}{\caption{\label{tab:inputs} Input parameters for
quark masses, CKM matrix element, and {\it effective} $B_{(s)}$ decay
constants, and for twist-3 3-parton LCDAs of the $K_{1A}$ and $K_{1B}$ states
\cite{PDG,Yang:2007zt}. The G-parity violating parameters are updated due to new values for $a_0^{\perp, K_{1A}}$ and $a_0^{\parallel, K_{1B}}$ given in Ref. \cite{Hatanaka:2008xj}.
}}}
\begin{center}
{\tabcolsep=0.989cm\begin{tabular} {|c c c c|}\hline\hline
\multicolumn{4}{|c|}{strange quark mass (GeV), pole $b$-quark mass (GeV), and couplings }
 \\
 \hline  ~~~$m_s(2\,\mbox{GeV})$  & ~~~$m_{b,pole}$ & ~~~$\alpha_s(1~{\rm GeV})$
 & $\alpha_s(2.2~{\rm GeV})$ \\
\hline
 $0.09\pm 0.01$ & $4.85\pm0.05$ & $0.495$ & $0.287$\\
\hline
\end{tabular}}
{\tabcolsep=1.08cm\begin{tabular}{|c | c| c |} \hline \multicolumn{3}{|c|}{the
CKM matrix element and the {\it effective} $B_{(s)}$ decay constants}
\\
\hline $~~~~~~~~~~|V_{ub}|$~~~~~~~~~~~~ &
 $f_{B}(\alpha_s=0)$  [MeV] &   $f_{B_s}(\alpha_s=0)$ [MeV] \\
\hline $(4.31\pm  0.30)\times 10^{-3}$ &  $145\pm 10$ &  $165\pm 10$ \\
\hline
\end{tabular}}
{\small{\tabcolsep=0.585cm\begin{tabular}{|c| c c c|} \hline
\multicolumn{4}{|c|}{G-parity conserving parameters of twist-3 3-parton LCDAs at the scale 2.2 GeV}
 \\
\hline
 &~~~~~~~$f^V_{3,^3P_1}$ [GeV$^2$]~~~~~& ~~~~~~~~~~ $\omega_{^3P_1}^V$~~~~~~~~
  & $f^A_{3,^3P_1}$ [GeV$^2$] \\
  \hline
  ~~~~~$a_1$~~~~ & $0.0036\pm 0.0018$ &  $-2.9  \pm 0.9$ & $0.0012\pm 0.0005$ \\
  ~~~~~$f_1$~~~~ & $0.0036\pm 0.0018$ &  $-2.8  \pm 0.9$ & $0.0012\pm 0.0005$ \\
  ~~~~~$f_8$~~~~ & $0.0035\pm 0.0018$ &  $-3.0  \pm 1.0$ & $0.0015\pm 0.0005$ \\
  ~~~~~$K_{1A}$~~~~ & $0.0034\pm 0.0018$&$-3.1  \pm 1.1$ & $0.0014\pm 0.0007$ \\
  \hline
  &~~~~~~~$f^A_{3,^1P_1}$ [GeV$^2$]~~~~~& ~~~~~~~~~~ $\omega_{^1P_1}^A$~~~~~~~~~~
  & $f^V_{3,^1P_1}$ [GeV$^2$] \\
  \hline
  ~~~~~$b_1$~~~~ & $-0.0036\pm 0.0014$ &  $-1.4  \pm 0.3$ & $0.0030\pm 0.0011$ \\
  ~~~~~$h_1$~~~~ & $-0.0033\pm 0.0014$ &  $-1.7  \pm 0.4$ & $0.0027\pm 0.0012$ \\
  ~~~~~$h_8$~~~~ & $-0.0035\pm 0.0014$ &  $-2.9  \pm 0.8$ & $0.0027\pm 0.0012$ \\
  ~~~~~$K_{1B}$~~~~& $-0.0041\pm 0.0018$& $-1.7  \pm 0.4$ & $0.0029\pm 0.0012$ \\
  \hline
\end{tabular}}
}
{\small{\tabcolsep=0.927cm\begin{tabular}{|ccc|} \hline
\multicolumn{3}{|c|}{G-parity violating parameters of twist-3 3-parton LCDAs of the $K_{1A}$ at $\mu=2.2$~GeV}
 \\
\hline
 ~~~~~~~~~~~~$\sigma_{K_{1A}}^V$~~~~~~~~~~~~~
&~~~~~~~~~ $\lambda_{K_{1A}}^A$~~~~~~~ & ~~~~~~$\sigma_{K_{1A}}^A$~~~~~~ \\
\hline
 $0.01\pm 0.04$ &  $-0.12  \pm 0.22$ & $-1.9\pm 1.1$\\
\hline
\end{tabular}}
}
{\small{\tabcolsep=0.92cm\begin{tabular}{|ccc|} \hline
\multicolumn{3}{|c|}{G-parity violating parameters of twist-3 3-parton LCDAs of the $K_{1B}$ at $\mu=2.2$~GeV}
 \\
\hline
 ~~~~~~~~~~~~ $\lambda_{K_{1B}}^V$~~~~~~~~~~~
&~~~~~~~~~~~$\sigma_{K_{1B}}^V$~~~~~~~~ & ~~~~~~~$\sigma_{K_{1B}}^A$~~~~~~ \\
\hline
 $-0.23 \pm 0.18$ & $ 1.3\pm 0.8$ &  $0.03\pm 0.03$\\
\hline\hline
\end{tabular}}}
\vskip0.8cm
\end{center}
\end{table}

\subsection{Numerical results for the form factors}\label{subsec:ff}

\begin{table}[ht!]
\caption{Parameters (in units of GeV$^2$) relevant to the excited-state
thresholds of the light-cone sum rules,  where $m_{m_b, pole}=4.85$~GeV is
used. Here parameters correspond to $q^2=0$.} \label{tab:sr-parameters}
\vspace{0.1cm}
\begin{ruledtabular}
\begin{tabular}{| rcccc  cccc |}
    & $B\to a_1$
    & $B\to f_1$
    & $B\to f_8$
    & $B\to K_{1A}$
    & $B\to b_1$
    & $B\to h_1$
    & $B\to h_8$
    & $B\to K_{1B}$
    \\
    \hline
$s_0$
   & 34.18
   & 34.17
   & 34.10
   & 34.14
   & 34.25
   & 34.90
   & 35.02
   & 34.23
    \\
$\delta_{V_2}$
   & 3.05
   & 3.15
   & 3.12
   & 3.55
   & 4.38
   & 7.48
   & 5.64
   & 3.75
   \\
$\delta_{V}$
   & 3.05
   & 3.15
   & 3.12
   & 3.55
   & 4.38
   & 7.48
   & 5.64
   & 3.75
   \\
$\delta_{A}$
   & 1.61
   & 1.55
   & 1.56
   & $-0.46$
   & 3.25
   & 3.42
   & 1.82
   & 2.55
\\
\end{tabular}
\end{ruledtabular}
\end{table}

We numerically analyze the light-cone sum rules for the transition form
factors, where the pole $b$ quark mass is adopted in the calculation. The
parameters appearing in the sum rules are evaluated at the factorization scale
$\mu_f=\sqrt{m_{B_q}^2-m_{b,pole}^2}$.

We find that, for $s_0 \simeq (34\sim 37)$ GeV$^2$, the $V_1$ sum rule can be
stable within the Borel mass range $6.0$~GeV$^2< M^2 < 12.0$~GeV$^2$. Therefore
we choose the Borel windows to be $6.0$~GeV$^2< M^2 < 12.0$~GeV$^2$,
$(6.0+\delta_{V_2})$~GeV$^2< M^2 < (12.0+\delta_{V_2})$~GeV$^2$, and
$(6.0+\delta_{A})$~GeV$^2< M^2 < (12.0+\delta_{A})$~GeV$^2$ for $V_1, V_2$, and
$A$, respectively, where the correction originating from higher resonance
states amounts to 8\% to 20\%. $V_0(0)$ equals to $V_3(0)$, where the latter
can be obtained from $V_1(0)$ and $V_2(0)$. As for $q^2\not= 0$, the Borel
windows for $V_3(q^2)-V_0(q^2)$ is $(6.0+\delta_{V})$~GeV$^2< M^2 <
(12.0+\delta_{V})$~GeV$^2$. The excited threshold $s_0$ is determined when the
most stable plateau of the $V_1$ sum rule result is obtained within the Borel
window. Using the same $s_0$, we can then determine $\delta_{V_1}, \delta_A$,
and $\delta_V$, so that the sum rule results for $V_2, A$, and $V_3-V_0$ are
stable within the Borel windows.  In Table \ref{tab:sr-parameters}, we show
that, for $q^2=0$ and $m_{b,pole}=4.85$~GeV, the corresponding $s_0$'s lie in
the interval $34 - 35$~GeV$^2$. The values of $\delta_{V_2, A, V}$,
corresponding to $m_{b,pole}=4.85$~GeV, are also collected in
Table~\ref{tab:sr-parameters}. In the present study, the excited thresholds
change slightly for larger $q^2$. However, for simplicity the values of $s_0$
and $\delta_{V_2, A, V}$ are chosen to be independent of $q^2$.

In the numerical analysis, we use the {\it effective} $B$ decay constant
$f_B(\alpha_s =0)=145\pm 10$~MeV, which is in agreement with the QCD sum rule
result without radiative corrections \cite{Khodjamirian:1998ji}. We have
checked that, using this value of $f_B$ and $m_b=4.85$~GeV in the light-cone
sum rules of $B\to \rho$ transition form factors of the same order of
$\alpha_s$ and $m_{\rho}/m_b$, we can get results: $A_1^{B\rho}(0)\simeq 0.23,
A_2^{B\rho}(0)\simeq 0.22$, and $V^{B\rho}(0) \simeq 0.32$, in good agreement
with that given in Ref.~\cite{Ball:2004rg}, where the radiative corrections are
included.  In the literature, it was found that the contributions due to
radiative corrections in the form factor sum rules can be canceled if one
adopts the $f_B$ sum rule result with the same order of $\alpha_s$-corrections
in the calculation \cite{Ball:1998kk,Ball:2004rg}. Therefore, radiative
corrections might be negligible in the present analysis.

Including the terms up to order of $m_A/m_b$ in the light-cone expansion, the
three-parton distribution amplitudes do not contribute directly to the sum
rules, but they enter the sum rules since $g_\perp^{(a)}$ and $g_\perp^{(v)}$
can be represented in terms of the leading-twist (two-parton) and twist-3
three-parton LCDAs. To estimate the theoretical uncertainties of the sum rule
results due to higher-twist effects, we put all parameters related to twist-3
three-parton LCDAs to be zero and find that the changes of the resulting form
factors are less than 3\%. We thus conclude that the higher-twist effects
might be negligible.

The form factors results in the light-cone sum rule calculation are exhibited
in Table \ref{tab:FF-BA}, where the momentum dependence is parameterized in
the three-parameter form:
 \begin{eqnarray} \label{eq:FFpara}
 F^{B_q A}(q^2)=\,{F^{B_q A}(0)\over 1-a(q^2/m_{B_q}^2)+b(q^2/m_{B_q}^2)^2}\,,
 \end{eqnarray}
with $F^{B_q A}\equiv V^{B_q A}_{0,1,2}$ or $A^{B_q A}$. For simplicity, we
do not show the theoretical errors for the parameters $a$ and $b$. As
$q^2\gtrsim 10$ GeV$^2$, the sum rule results become less stable. To get
reliable estimate for the $q^2$-dependence of the form factors, the present
results are fitted in the range $0\leq q^2 \leq 6$~GeV$^2$. The theoretical
errors for $F(0)$ are due to variation of the Borel mass, the Gegenbauer
moments, the decay constants, the strange quark mass, and the pole $b$ quark
mass, which are then added in quadrature. The errors are dominated by
variation of the pole $b$ quark mass. It should be stressed that in the
convention of the present work, the decay constants of $1^1P_1$ and $1^3P_1$
axial-vector mesons are of the same sign, so that the form factors for $B\to
1^1P_1$ and $B\to 1^3P_1$ transitions have opposite signs. The sign convention
is the other way around in the light-front quark model \cite{CCH} and
perturbative QCD \cite{pQCD} calculations.

For the numerical analysis of $B_s\to $ form factors, we adopt the {\it
effective} decay constant $f_{B_s}(\alpha_s =0) \simeq 1.14\times
f_{B_d}(\alpha_s =0) \simeq 165\pm 11$~MeV, which is estimated by using the
relevant QCD sum rule result. Finally, we obtain the relations (with $F\equiv
V_{0,1,2}$ or $A$):
\begin{eqnarray}
& & \frac{F^{B K_{1A}}(q^2)}{F^{B_s K_{1A}}(q^2)}\simeq
\frac{F^{Bf_1}(q^2)}{F^{B_s f_1}(q^2)}\simeq
-\frac{2 F^{Bf_8}(q^2)}{F^{B_s f_8}(q^2)} \nonumber\\
& \simeq & \frac{F^{B K_{1B}}(q^2)}{F^{B_s K_{1B}}(q^2)}\simeq
\frac{F^{Bh_1}(q^2)}{F^{B_s h_1}(q^2)}\simeq
-\frac{2 F^{Bh_8}(q^2)}{F^{B_s h_8}(q^2)} \simeq 1.14.
\end{eqnarray}
In Table \ref{tab:FF-BsA}, we show the form factor results at the maximum
recoil (i.e., at $q^2=0$).

\begin{table}[t]
\caption{Form factors for $B_{u,d}\to a_1,b_1,K_{1A},K_{1B}, f_1, f_8, h_1,
h_8$ transitions obtained in the light-cone sum rule calculation are fitted to
the 3-parameter form in Eq. (\ref{eq:FFpara}). Here, because the decay
constants, $f_{^3P_1}$ and $f_{^1P_1}^\perp$, which are defined in Eqs.
(\ref{eq:decay-const1.1}) and (\ref{eq:decay-const1.2}), are of the same sign,
the form factors for $B\to 1^1P_1$ and $B\to 1^3P_1$ transitions have opposite
signs.} \label{tab:FF-BA}
\begin{ruledtabular}
\begin{tabular}{| c c c c || c c c c |}
~~~$F$~~~~~
    & $F(0)$~~~~~
    &$a$~~~~~
    & $b$~~~~~~
& ~~~ $F$~~~~~
    & $F(0)$~~~~~
    & $a$~~~~~
    & $b$~~~~~~
 \\
    \hline
$V_1^{Ba_1}$
    & $0.37\pm0.07$
    & ~~$0.645$
    & $0.250$
&$V_1^{Bb_1}$
    & $-0.20\pm 0.04$
    & ~~$0.748$
    & ~~$0.063$
    \\
$V_2^{Ba_1}$
    & $0.42\pm0.08$
    & $1.48$
    & $1.00~~$
&$V_2^{Bb_1}$
    & $-0.09\pm0.02$
    & ~~$0.539$
    & $1.76$
    \\
$V_0^{Ba_1}$
    & $0.30\pm 0.05$
    & $1.77$
    & $0.926$
&$V_0^{Bb_1}$
    & $-0.39\pm0.07$
    & $1.22$
    & ~~$0.426$
    \\
$A^{Ba_1}$
    & $0.48\pm0.09$
    & $1.64$
    & $0.986$
&$A^{Bb_1}$
    & $-0.25\pm0.05$
    & $1.69$
    & ~~$0.910$
    \\
$V_1^{BK_{1A}}$
    & $0.34\pm0.07$
    & ~~$0.635$
    & $0.211$
&$V_1^{BK_{1B}}$
    & $-0.29^{+0.08}_{-0.05}$
    & ~~$0.729$
    & ~~$0.074$
    \\
$V_2^{BK_{1A}}$
    & $0.41\pm 0.08$
    & $1.51$
    & $1.18~~$
&$V_2^{BK_{1B}}$
    & $-0.17^{+0.05}_{-0.03}$
    & ~~$0.919$
    & ~~$0.855$
    \\
$V_0^{BK_{1A}}$
    & $0.22\pm0.04$
    & $2.40$
    & $1.78~~$
&$V_0^{BK_{1B}}$
    & $-0.45^{+0.12}_{-0.08}$
    & $1.34$
    & ~~$0.690$
    \\
$A^{BK_{1A}}$
    & $0.45\pm0.09$
    & $1.60$
    & $0.974$
&$A^{BK_{1B}}$
    & $-0.37^{+0.10}_{-0.06}$
    & $1.72$
    & ~~$0.912$
    \\
$V_1^{Bf_1}$
    & $0.23\pm0.04$
    & ~~$0.640$
    & $0.153$
&$V_1^{Bh_1}$
    & $-0.13\pm 0.03$
    & ~~$0.612$
    & ~~$0.078$
    \\
$V_2^{Bf_1}$
    & $0.26\pm0.05$
    & $1.47$
    & $0.956$
&$V_2^{Bh_1}$
    & $-0.07\pm0.02$
    & ~~$0.500$
    & $1.63$
    \\
$V_0^{Bf_1}$
    & $0.18\pm 0.03$
    & $1.81$
    & $0.880$
&$V_0^{Bh_1}$
    & $-0.24\pm0.04$
    & $1.16$
    & ~~$0.294$
    \\
$A^{Bf_1}$
    & $0.30\pm0.05$
    & $1.63$
    & $0.900$
&$A^{Bh_1}$
    & $-0.17\pm0.03$
    & $1.54$
    & ~~$0.848$
    \\
$V_1^{Bf_8}$
    & $0.16\pm0.03$
    & ~~$0.644$
    & $0.209$
&$V_1^{Bh_8}$
    & $-0.11\pm 0.02$
    & ~~$0.623$
    & ~~$0.094$
    \\
$V_2^{Bf_8}$
    & $0.19\pm0.03$
    & $1.49$
    & $1.09~~$
&$V_2^{Bh_8}$
    & $-0.06\pm0.01$
    & ~~$0.529$
    & $1.53$
    \\
$V_0^{Bf_8}$
    & $0.12\pm 0.02$
    & $1.84$
    & $0.749$
&$V_0^{Bh_8}$
    & $-0.18\pm0.03$
    & $1.22$
    & ~~$0.609$
    \\
$A^{Bf_8}$
    & $0.22\pm0.04$
    & $1.64$
    & $0.919$
&$A^{Bh_8}$
    & $-0.13\pm0.02$
    & $1.56$
    & ~~$0.827$
    \\
\end{tabular}
\end{ruledtabular}
\end{table}
%

\begin{table}[t]
\caption{Form factors for $B_s \to K_{1A},K_{1B}, f_1, f_8, h_1, h_8$
transitions obtained in the light-cone sum rule calculation are fitted to the
3-parameter form in Eq. (\ref{eq:FFpara}). Here, because the decay constants,
$f_{^3P_1}$ and $f_{^1P_1}^\perp$, are of the same sign, the form factors for
$B_s\to 1^1P_1$ and $B_s\to 1^3P_1$ transitions have opposite signs.}
\label{tab:FF-BsA}
\begin{ruledtabular}
\begin{tabular}{| c r c c || c l c c |}
~~~$F$~~~~~
    & $F(0)$~~~~~
    &$a$~~~~~
    & $b$~~~~~~
& ~~~ $F$~~~~~
    & $F(0)$~~~~~
    & $a$~~~~~
    & $b$~~~~~~
 \\
    \hline
$V_1^{B_sK_{1A}}$
    & $0.30\pm0.06$
    & ~~$0.635$
    & $0.211$
&$V_1^{B_sK_{1B}}$
    & $-0.25^{+0.07}_{-0.04}$
    & ~~$0.729$
    & ~~$0.074$
    \\
$V_2^{B_sK_{1A}}$
    & $0.36\pm 0.07$
    & $1.51$
    & $1.18~~$
&$V_2^{B_s K_{1B}}$
    & $-0.15^{+0.04}_{-0.03}$
    & ~~$0.919$
    & ~~$0.855$
    \\
$V_0^{B_s K_{1A}}$
    & $0.19\pm0.04$
    & $2.40$
    & $1.78~~$
&$V_0^{B_s K_{1B}}$
    & $-0.40^{+0.11}_{-0.07}$
    & $1.34$
    & ~~$0.690$
    \\
$A^{B_s K_{1A}}$
    & $0.40\pm0.08$
    & $1.60$
    & $0.974$
&$A^{B_s K_{1B}}$
    & $-0.33^{+0.09}_{-0.05}$
    & $1.72$
    & ~~$0.912$
    \\
$V_1^{B_s f_1}$
    & $0.20\pm0.04$
    & ~~$0.640$
    & $0.153$
&$V_1^{B_s h_1}$
    & $-0.11\pm 0.03$
    & ~~$0.612$
    & ~~$0.078$
    \\
$V_2^{B_s f_1}$
    & $0.23\pm0.04$
    & $1.47$
    & $0.956$
&$V_2^{B_s h_1}$
    & $-0.06\pm0.02$
    & ~~$0.500$
    & $1.63$
    \\
$V_0^{B_s f_1}$
    & $0.16\pm 0.03$
    & $1.81$
    & $0.880$
&$V_0^{B_s h_1}$
    & $-0.21\pm0.04$
    & $1.16$
    & ~~$0.294$
    \\
$A^{B_s f_1}$
    & $0.26\pm0.04$
    & $1.63$
    & $0.900$
&$A^{B_s h_1}$
    & $-0.15\pm0.03$
    & $1.54$
    & ~~$0.848$
    \\
$V_1^{B_s f_8}$
    & $-0.28\pm0.05$
    & ~~$0.644$
    & $0.209$
&$V_1^{B_s h_8}$
    & ~~$0.19\pm 0.04$
    & ~~$0.623$
    & ~~$0.094$
    \\
$V_2^{B_s f_8}$
    & $-0.33\pm0.05$
    & $1.49$
    & $1.09~~$
&$V_2^{B_s h_8}$
    & ~~$0.11\pm0.02$
    & ~~$0.529$
    & $1.53$
    \\
$V_0^{B_s f_8}$
    & $-0.21\pm 0.04$
    & $1.84$
    & $0.749$
&$V_0^{B_s h_8}$
    & ~~$0.32\pm0.05$
    & $1.22$
    & ~~$0.609$
    \\
$A^{B_s f_8}$
    & $-0.39\pm0.07$
    & $1.64$
    & $0.919$
&$A^{B_s h_8}$
    & ~~$0.23\pm0.04$
    & $1.56$
    & ~~$0.827$
    \\
\end{tabular}
\end{ruledtabular}
\vskip0.7cm
\end{table}

\subsection{Branching ratios}

Our results for the semileptonic decay rates $B_{u,d,s}\to A e \bar\nu_e$ are
listed in Table~\ref{tab:rates}. Most branching ratios are of order $10^{-4}$.
For $B_{u,d}$ decays involving the $a_1$ or $f_1(1285)$ we obtain $
\Gamma_L/\Gamma_T \simeq 0.6$, whereas for decays containing $1^1P_1$ mesons
we find that $\Gamma_L/\Gamma_T$ is close to 2. In short, the polarization
fractions follow the relations: $\Gamma_-> \Gamma_L \gg \Gamma_+$ for the
former, and $\Gamma_L> \Gamma_- \gg \Gamma_+$ for the latter.  On the other
hand,  we have $\Gamma_L/\Gamma_T \sim 0.6-0.7, 1.1, 1.4-1.8$ for the
semileptonic $B_s$ decays involving the $f_1, h_1$ and $K_1$, respectively.
These results are sensitive to the values of form factors. Moreover, we have
the salient patterns
\begin{eqnarray}
{\cal B}^{u}[a_1(1260)]>{\cal B}^{u}[b_1(1235)]>{\cal B}^{u}[f_1(1285)]
 > {\cal B}^{u}[h_1(1170)] \gg {\cal B}^{u}[f_1(1420)]\gtrsim {\cal B}^{u}[h_1(1380)],
 ~~~~\\
{\cal B}^{s}[K_1(1270)]\gtrsim {\cal B}^{s}[K_1(1400)]>{\cal B}^{s}[f_1(1420)]
 > {\cal B}^{s}[h_1(1380)] \gg {\cal B}^{s}[f_1(1285)]\gtrsim {\cal
B}^{s}[h_1(1170)], ~~~~
\end{eqnarray}
where ${\cal B}^{q}[a_1(1260)]\equiv {\cal B}(B_{q}\to A e \bar\nu_e)$. These
patterns are sensitive to the mixing angles and thus can offer deeper insights
for the quark contents of the $P$-wave mesons.

%
\begin{table}[ht!]
\caption{Decay rates of $B_{u,d,s}\to A e \bar\nu_e$ obtained in this work,
where $\Gamma_{\pm,L}$ are in units of $10^{6}s^{-1}$, and the branching
ratios are in units of $10^{-4}$. $\Gamma_L$ stands for the portion of the
rate with a longitudinal polarization $A$, $\Gamma_+$ with a positive helicity
$A$, and $\Gamma_-$ with a negative helicity $A$. $\Gamma_T=\Gamma_+ +
\Gamma_-$. Here we use  $\theta_{K_1}=-(34 \pm 13)^\circ$,
$\theta_{^3P_1}=(23.6^{+17.0}_{-23.6})^\circ$, and
$\theta_{^1P_1}=(28.1^{+~9.8}_{-17.2})^\circ$. The first error comes from the
variation of form factors, and the second from the mixing angles.}
\label{tab:rates} {\footnotesize \begin{ruledtabular}
\begin{tabular}{|c| clll ll|}
 $A$
    & $\Gamma_+$
    & ~~~~$\Gamma_-$
    & ~~~~$\Gamma_L$
    & $\Gamma_L/\Gamma_T$
    & ${\cal B}(\bar B^0\to A^+ e^- \bar\nu_e)$
    & ${\cal B}(B^-\to A^0 e^- \bar\nu_e)$
    \\
    \hline
$a_1(1260)$
   & $7.6^{+3.2}_{-2.6} $~~~~~
   & $\!\!\! 115^{+48}_{-39} $
   & $74.8^{+30.7}_{-25.4} $
   & $0.61^{+0.00}_{-0.00}$
   & $3.02^{+1.03}_{-1.03} $
   & $3.24^{+1.33}_{-1.13} $
    \\
$f_1(1285)$
   & $4.2^{+1.7+0.0}_{-1.4-1.1} $
   & $62.1^{+23.6+~2.1}_{-19.8-18.9} $
   & $40.4^{+14.9+~0.5}_{-12.5-11.3} $
   & $0.61^{+0.01+0.02}_{-0.00-0.01} $
   & $1.63^{+0.60+0.04}_{-0.51-0.48} $
   & $1.75^{+0.65+0.04}_{-0.55-0.52} $
   \\
$f_1(1420)$
   & $0.1^{+0.1+1.2}_{-0.0-0.0} $
   & ~\,$2.2^{+1.0+15.7}_{-0.8-~1.7} $
   & ~\,$1.1^{+0.9+10.3}_{-1.0-~0.5} $
   & $0.48^{+0.10+0.60}_{-0.14-0.00}$
   & $0.05^{+0.03+0.42}_{-0.02-0.03} $
   & $0.06^{+0.03+0.44}_{-0.03-0.04} $
   \\
$b_1(1235)$
   & $3.3^{+1.4}_{-1.2}$~~~~~
   & $37.7^{+16.6}_{-13.5}$
   & $85.5^{+36.5}_{-30.2}$
   & $2.08^{+0.03}_{-0.01}$
   & $1.93^{+0.84}_{-0.68}$
   & $2.07^{+0.90}_{-0.73}$
   \\
$h_1(1170)$
   & $1.9^{+1.2+0.1}_{-0.9-0.4}$
   & $24.7^{+10.8+1.0}_{-~8.8-4.8}$
   & $54.7^{+24.8+~2.4}_{-20.2-11.2}$
   & $2.05^{+0.01+0.01}_{-0.01-0.01}$
   & $1.24^{+0.57+0.06}_{-0.45-0.25}$
   & $1.33^{+0.60+0.06}_{-0.49-0.27}$
   \\
$h_1(1380)$
   & $0.1^{+0.2+0.5}_{-0.0-0.1}$
   & ~\,$0.8^{+0.1+3.7}_{-0.2-0.8}$
   & ~~$1.8^{+0.5+7.8}_{-0.5-1.7}$
   & $1.97^{+0.03+0.12}_{-0.14-0.07}$
   & $0.04^{+0.01+0.18}_{-0.01-0.00}$
   & $0.04^{+0.02+0.20}_{-0.01-0.00}$
   \\
   \hline
 $A$
    & $\Gamma_+$
    & ~~~~$\Gamma_-$~~
    & ~~~~$\Gamma_L$
    & $\Gamma_L/\Gamma_T$
    & ${\cal B}(B_s\to A^+ e^- \bar\nu_e)$
    &
    \\
    \hline
$f_1(1285)$
   & $0.4^{+1.0+2.1}_{-0.2-0.3} $
   & \, $4.4^{+9.0+32.8}_{-4.2-~3.3} $
   & \, $3.3^{+8.2+20.1}_{-3.2-~2.9} $
   & $0.68^{+0.07+0.00}_{-0.42-0.36} $
   & $0.12^{+0.27+0.00}_{-0.12-0.10} $
   &
   \\
$f_1(1420)$
   & $6.2^{+0.9+0.4}_{-0.8-2.1} $
   & $90.8^{+17.9+~2.8}_{-16.3-27.3} $
   & $57.6^{+12.3+~2.6}_{-11.2-18.3} $
   & $0.59^{+0.01+0.02}_{-0.01-0.01}$
   & $2.27^{+0.45+0.08}_{-0.42-0.70} $
   &
   \\
$h_1(1170)$
   & $0.1^{+0.1+0.2}_{-0.0-0.0}$
   & \, $0.1^{+0.2+5.7}_{-0.1-~0.0}$
   & \, $0.2^{+0.3+11.6}_{-0.2-~0.0}$
   & $1.37^{+0.02+0.71}_{-0.48-0.00}$
   & $0.01^{+0.00+0.25}_{-0.00-0.00}$
   &
   \\
$h_1(1380)$
   & $3.7^{+2.2+0.0}_{-1.7-0.3}$
   & $37.6^{+17.4+0.0}_{-14.1-4.4}$
   & $74.2^{+37.7+0.0}_{-30.0-8.2}$
   & $1.80^{+0.04+0.00}_{-0.06-0.01}$
   & $1.69^{+0.84+0.00}_{-0.67-0.19}$
   &
   \\
$K_1(1270)$
   & $9.2^{+3.7+0.6}_{-4.0-1.4} $
   & $\!\!\! 141^{+54+~9}_{-61-22} $
   & $\!\!\! 159^{+57+~0}_{-72-10} $
   & $1.06^{+0.00+0.15}_{-0.05-0.12}$
   & $4.53^{+1.67+0.00}_{-2.00-0.44} $
   &
   \\
$K_1(1400)$
   & $9.4^{+3.4+0.6}_{-4.1-1.5} $
   & $\!\!\! 119^{+45+~8}_{-51-19} $
   & $135^{+49+0}_{-61-7} $
   & $1.05^{+0.00+0.13}_{-0.04-0.11}$
   & $3.86^{+1.43+0.03}_{-1.70-0.40} $
   &
    \\
   \end{tabular}
\end{ruledtabular}}
\vskip0.5cm
\end{table}

\section{Summary}\label{sec:summary}

We have calculated the vector and axial-vector form factors of $B$ decays into
$P$-wave axial-vector mesons in the light-cone sum rule approach. Owing to the
$G$-parity, the chiral-even two-parton light-cone distribution amplitudes of
the $1^3P_1$ and $1^1P_1$ mesons are respectively symmetric and antisymmetric
under the exchange of quark and anti-quark momentum fractions in the SU(3)
limit. For chiral-odd light-cone distribution amplitudes, it is the other way
around. The sum rule results for form factors are sensitive to the light-cone
distribution amplitudes of the axial-vector mesons. To extract the relevant
form factors, the polarization of the axial-vector meson is chosen to be
transversely polarized in the light-cone sum rule calculation. For the
resulting sum rules, we have included the terms up to order of $m_A/m_b$ in the
light-cone expansion. The numerical impact of $1/m_b$ corrections is under
control. As discussed in Sec.~\ref{sec:lcsr}, it should be stressed that the
expansion parameter in the light-cone sum rules is $m_A/m_b$, instead of twist.
In the present study, because the decay constants, $f_{^3P_1}$ and
$f_{^1P_1}^\perp$, which are defined in Eqs. (\ref{eq:decay-const1.1}) and
(\ref{eq:decay-const1.2}), are of the same sign, the form factors for $B\to
1^1P_1$ and $B\to 1^3P_1$ transitions have opposite signs. The sum rule results
could be improved in the future by including ${\cal O}(\alpha_s)$ corrections
to the sum rules and by improving the input parameters describing the
light-meson distribution amplitudes, for instance from lattice calculations. We
have presented the results for the semileptonic decay rates of $B_{u,d,s} \to A
e \bar\nu_e$. Theses will allow further tests of our form factor results and of
the mixing angles in the future.

\section*{Acknowledgments}

This research was supported in part by the
National Science Council of R.O.C. under Grant No. NSC96-2112-M-033-004-MY3.

%
\appendix
\section{Two-parton distribution amplitudes}\label{app:2da-def}

The chiral-even LCDAs are given by
\begin{eqnarray}
  \langle A(P,\lambda)|\bar q_1(x) \gamma_\mu \gamma_5 q_2(0)|0\rangle
  && = if_A m_A \, \int_0^1
      du \,  e^{i u \, p x}
   \left\{p_\mu \,
    \frac{\epsilon^{*(\lambda)} x}{p x} \, \Phi_\parallel(u)
         +\varepsilon_{\perp\mu}^{*(\lambda)} \,
    g_\perp^{(a)}(u) \right. \nonumber\\
    &&\ \ \left. - \frac{1}{2}x_{\mu}
\frac{\epsilon^{*(\lambda)} x }{(p  x)^{2}} m_{A}^{2} g_{3}(u)
 \right\},
                                 \label{eq:evendef1} \\
  \langle A(P,\lambda)|\bar q_1(x) \gamma_\mu
  q_2(0)|0\rangle
  && = - i f_A m_A
\,\epsilon_{\mu\nu\rho\sigma} \,
      \epsilon^{*\nu}_{(\lambda)} p^{\rho} x^\sigma \, \int_0^1 du \,  e^{i u \, p x} \,
      \Bigg( \frac{g_\perp^{(v)}(u)}{4} \Bigg),\ \ \ \ \ \ \
       \label{eq:evendef2}
\end{eqnarray}
where $u$ and $\bar u\equiv 1-u$ are the momentum fractions carried by the
$q_1$ and $\bar q_2$ quarks, respectively, in the axial-vector meson. The
chiral-odd LCDAs are defined by
\begin{eqnarray}
  &&\langle A(P,\lambda)|\bar q_1(x) \sigma_{\mu\nu}\gamma_5 q_2(0)
            |0\rangle
  =  f_A^{\perp} \,\int_0^1 du \, e^{i u \, p x} \,
\Bigg\{(\varepsilon^{*(\lambda)}_{\perp\mu} p_{\nu} -
  \varepsilon_{\perp\nu}^{*(\lambda)}  p_{\mu})
  \Phi_\perp(u)\nonumber\\
&& \hspace*{+5cm}
  + \,\frac{m_A^2\,\epsilon^{*(\lambda)} x}{(p x)^2} \,
   (p_\mu x_\nu -
    p_\nu  x_\mu) \, h_\parallel^{(t)}(u)\nonumber\\
 && \hspace*{+5cm} + \frac{1}{2}
(\varepsilon^{*(\lambda)}_{\perp \mu} x_\nu
-\varepsilon^{*(\lambda)}_{\perp \nu} x_\mu) \frac{m_{A}^{2}}{p x}  h_{3}(u)
\Bigg\},\label{eq:odddef1}\\
&&\langle A(P,\lambda)|\bar q_1(x) \gamma_5 q_2(0) |0\rangle
  =  f_A^\perp
 m_{A}^2 (\epsilon^{*(\lambda)} x)\,\int_0^1 du \, e^{i u \, p x}  \,
\Bigg(\frac{h_\parallel^{(p)}(u)}{2} \Bigg),\ \ \ \ \ \ \
\label{eq:odddef2}
\end{eqnarray}
where $p_\mu=P_\mu -m_A^2 x_\mu/(2Px)$ and $x^2 =0$. Here, $\Phi_\parallel, \Phi_\perp$ are
of twist-2, $g_\perp^{(a)}, g_\perp^{(v)}, h_\parallel^{(t)},
h_\parallel^{(p)}$ of twist-3, and $g_3, h_3$ of twist-4. Note that in the
definitions of LCDAs, the longitudinal and transverse {\it projections} of
polarization vectors $\epsilon^{*(\lambda)}_\mu$ along the $x$-direction for
the axial-vector meson are given by
\begin{eqnarray}\label{eq:polprojectiors}
 && \varepsilon^{*(\lambda)}_{\parallel\, \mu} \equiv
     \frac{\epsilon^{*(\lambda)} x}{p x} \left(
      p_\mu-\frac{m_{A}^2}{2 p x} \,x_\mu\right), \qquad
 \varepsilon^{*(\lambda)}_{\perp\, \mu}
        = \epsilon^{*(\lambda)}_\mu -\varepsilon^{*(\lambda)}_{\parallel\,
        \mu}\,.
\end{eqnarray}
One should distinguish the above projections from the {\it exactly}
longitudinal ($\epsilon^{*(0)\mu}$) and transverse
($\epsilon^{*(\lambda)\mu}_\perp$) polarization vectors of the axial-vector
meson, given in Eq.~(\ref{eq:exact-polarization}) where the results are
independent of the coordinate variable $x$.

 In SU(3) limit, due to $G$-parity,
$\Phi_\parallel, g_\perp^{(a)}$, $g_\perp^{(v)}$, and $g_3$ are
symmetric (antisymmetric) under the replacement $u\to 1-u$ for the
$1^3P_1$ ($1^1P_1$) states, whereas $\Phi_\perp, h_\parallel^{(t)}$,
$h_\parallel^{(p)}$, and $h_3$ are antisymmetric (symmetric). In
other words, in the SU(3) limit it follows that
\begin{equation}
 \int_0^1 du \Phi_\perp(u)=\int_0^1 du h_\parallel^{(t)}(u)
 =\int_0^1 du h_\parallel^{(p)}(u)=\int_0^1 du h_3(u)=0
 \end{equation}
for $1^3P_1$ states, but becomes
\begin{equation}
 \int_0^1 du \Phi_\parallel(u)=\int_0^1 du g_\perp^{(a)}(u)
 =\int_0^1 dug_\perp^{(v)}(u)=\int_0^1 du g_3(u)=0
 \end{equation}
for $1^1P_1$ states. The above integrals are not zero if
$m_{q_1}\neq m_{q_2}$, and the detailed discussions can be found in
Ref.~\cite{Yang:2007zt}.  For
convenience, we therefore normalize the distribution amplitudes of
the $1^3P_1$ and $1^1P_1$ states to be subject to
 \begin{equation}
 \int_0^1 du\Phi_\parallel^{^3P_1}(u)=1 \ \ \
 \hbox{and} \ \ \int_0^1 du\Phi_\perp^{^1P_1}(u)=1.
 \end{equation}
We set $f_{^3\! P_1}^\perp=f_{^3\! P_1}$ and $f_{^1\! P_1}=f_{^1\!
P_1}^\perp(\mu=1~{\rm GeV})$ in the study, such that we have
 \begin{eqnarray}
 &&\langle 1^3P_1(P,\lambda)|
  \bar q_1(0) \sigma_{\mu\nu}\gamma_5 q_2(0)
   |0\rangle
  =  f_{^3P_1} a_0^{\perp,^3P_1} \,
(\epsilon^{*(\lambda)}_{\mu} P_{\nu} - \epsilon_{\nu}^{*(\lambda)}
P_{\mu}),
 \\
 &&  \langle 1^1P_1(P,\lambda)|\bar q_1(0) \gamma_\mu \gamma_5 q_2(0)|0\rangle
   = if_{^1P_1}^{\perp}(1~{\rm GeV}) a_0^{\parallel,^1P_1} \, m_{^1P_1} \,
    \epsilon^{*(\lambda)}_\mu
   \,,
 \end{eqnarray}
where $a_0^{\perp,^3P_1}$ and  $a_0^{\parallel,^1P_1}$ are the
Gegenbauer zeroth moments and vanish in the SU(3) limit.

We take into account the approximate forms of twist-2 distributions
for the $1^3P_1$ mesons to be \cite{Yang:2007zt}
\begin{eqnarray}
\Phi_\parallel(u) & = & 6 u \bar u \left[ 1 + 3 a_1^\parallel\, \xi +
a_2^\parallel\, \frac{3}{2} ( 5\xi^2  - 1 )
 \right], \label{eq:lcda-3p1-t2-1}\\
 \Phi_\perp(u) & = & 6 u \bar u \left[ a_0^\perp + 3 a_1^\perp\, \xi +
a_2^\perp\, \frac{3}{2} ( 5\xi^2  - 1 ) \right], \label{eq:lcda-3p1-t2-2}
\end{eqnarray}
and for the $1^1P_1$ mesons to be
\begin{eqnarray}
 \Phi_\parallel(u) & = & 6 u \bar u \left[ a_0^\parallel + 3
a_1^\parallel\, \xi +
a_2^\parallel\, \frac{3}{2} ( 5\xi^2  - 1 ) \right], \label{eq:lcda-1p1-t2-1}\\
\Phi_\perp(u) & = & 6 u \bar u \left[ 1 + 3 a_1^\perp\, \xi +
a_2^\perp\, \frac{3}{2} ( 5\xi^2  - 1 ) \right],
\label{eq:lcda-1p1-t2-2}
\end{eqnarray}
where $\xi=2u-1$.

For the relevant two-parton twist-3 chiral-even LCDAs, we take the approximate
expressions up to conformal spin $9/2$ and ${\cal O}(m_s)$
\cite{Yang:2007zt}:
\begin{eqnarray}
 g_\perp^{(a)}(u) & = &  \frac{3}{4}(1+\xi^2)
+ \frac{3}{2}\, a_1^\parallel\, \xi^3
 + \left(\frac{3}{7} \,
a_2^\parallel + 5 \zeta_{3,^3P_1}^V \right) \left(3\xi^2-1\right)
 \nonumber\\
& & {}+ \left( \frac{9}{112}\, a_2^\parallel + \frac{105}{16}\,
 \zeta_{3,^3P_1}^A - \frac{15}{64}\, \zeta_{3,^3P_1}^V \omega_{^3P_1}^V
 \right) \left( 35\xi^4 - 30 \xi^2 + 3\right) \nonumber\\
 & &
 + 5\Bigg[ \frac{21}{4}\zeta_{3,^3P_1}^V \sigma_{^3P_1}^V
  + \zeta_{3,^3P_1}^A \bigg(\lambda_{^3P_1}^A -\frac{3}{16}
 \sigma_{^3P_1}^A\Bigg) \Bigg]\xi(5\xi^2-3)
 \nonumber\\
& & {}-\frac{9}{2} {a}_1^\perp
\,\widetilde{\delta}_+\,\left(\frac{3}{2}+\frac{3}{2}\xi^2+\ln u
 +\ln\bar{u}\right) - \frac{9}{2} {a}_1^\perp\,\widetilde{\delta}_-\, (
3\xi + \ln\bar{u} - \ln u), \label{eq:ga-3p1}\\
g_\perp^{(v)}(u) & = & 6 u \bar u \Bigg\{ 1 +
 \Bigg(a_1^\parallel + \frac{20}{3} \zeta_{3,^3P_1}^A
 \lambda_{^3P_1}^A\Bigg) \xi\nonumber\\
 && + \Bigg[\frac{1}{4}a_2^\parallel + \frac{5}{3}\,
 \zeta^V_{3,^3P_1} \left(1-\frac{3}{16}\, \omega^V_{^3P_1}\right)
 +\frac{35}{4} \zeta^A_{3,^3P_1}\Bigg] (5\xi^2-1) \nonumber\\
 &&+ \frac{35}{4}\Bigg(\zeta_{3,^3P_1}^V
 \sigma_{^3P_1}^V -\frac{1}{28}\zeta_{3,^3P_1}^A
 \sigma_{^3P_1}^A \Bigg) \xi(7\xi^2-3) \Bigg\}\nonumber\\
& & {} -18 \, a_1^\perp\widetilde{\delta}_+ \,  (3u \bar{u} +
\bar{u} \ln \bar{u} + u \ln u ) - 18\,
a_1^\perp\widetilde{\delta}_- \,  (u \bar u\xi + \bar{u} \ln \bar{u} -
u \ln u),
 \label{eq:gv-3p1}
 \end{eqnarray}
for the $1^3P_1$ states, and
\begin{eqnarray}
 g_\perp^{(a)}(u) & = & \frac{3}{4} a_0^\parallel (1+\xi^2)
+ \frac{3}{2}\, a_1^\parallel\, \xi^3
 + 5\left[\frac{21}{4} \,\zeta_{3,^1P_1}^V
 + \zeta_{3,^1P_1}^A \Bigg(1-\frac{3}{16}\omega_{^1P_1}^A\Bigg)\right]
 \xi\left(5\xi^2-3\right)
 \nonumber\\
& & {}+ \frac{3}{16}\, a_2^\parallel \left(15\xi^4 -6 \xi^2 -1\right)
 + 5\, \zeta^V_{3,^1P_1}\lambda^V_{^1P_1}\left(3\xi^2 -1\right)
 \nonumber\\
& & {}+ \frac{105}{16}\left(\zeta^A_{3,^1P_1}\sigma^A_{^1P_1}
-\frac{1}{28} \zeta^V_{^1P_1}\sigma^V_{^1P_1}\right)
 \left(35\xi^4 -30 \xi^2 +3\right)\nonumber\\
 & & {}-15 {a}_2^\perp \bigg[ \widetilde{\delta}_+ \xi^3 +
 \frac{1}{2}\widetilde{\delta}_-(3\xi^2-1) \bigg] \nonumber\\
& & {}
  -\frac{3}{2}\,\bigg[\widetilde{\delta}_+\, ( 2 \xi + \ln\bar{u} -\ln u)
 +\, \widetilde{\delta}_-\,(2+\ln u + \ln\bar{u})\bigg](1+6a_2^\perp)
 ,\label{eq:ga-1p1}\\
g_\perp^{(v)}(u) & = & 6 u \bar u \Bigg\{ a_0^\parallel +
a_1^\parallel \xi +
 \Bigg[\frac{1}{4}a_2^\parallel
  +\frac{5}{3} \zeta^V_{3,^1P_1}
  \Bigg(\lambda^V_{^1P_1} -\frac{3}{16} \sigma^V_{^1P_1}\Bigg)
  +\frac{35}{4} \zeta^A_{3,^1P_1}\sigma^A_{^1P_1}\Bigg](5\xi^2-1) \nonumber\\
  & & {}  + \frac{20}{3}\,  \xi
 \left[\zeta^A_{3, ^1P_1}
 + \frac{21}{16}
 \Bigg(\zeta^V_{3,^1P_1}- \frac{1}{28}\, \zeta^A_{3,^1P_1}\omega^A_{^1P_1}
  \Bigg)
 (7\xi^2-3)\right]\nonumber\\
 & & {} -5\, a_2^\perp [2\widetilde\delta_+ \xi + \widetilde\delta_- (1+\xi^2)]
 \Bigg\}\nonumber\\
 & & {} - 6 \bigg[\, \widetilde{\delta}_+ \, (\bar{u} \ln\bar{u} -u\ln u )
  +\, \widetilde{\delta}_- \, (2u \bar{u} + \bar{u} \ln \bar{u} + u \ln u)\bigg]
  (1+6 a_2^\perp) ,
 \label{eq:gv-1p1}
\end{eqnarray}
for the $1^1P_1$ states, where
\begin{equation}
\widetilde{\delta}_\pm  ={f_{A}^{\perp}\over f_{A}}{m_{q_2} \pm m_{q_1}
\over m_{A}},\qquad \zeta_{3,A}^{V(A)} = \frac{f^{V(A)}_{3A}}{f_{A} m_{A}}.
\label{eq:parameters3}
\end{equation}

\section{An alternative definition of form factors}\label{app:formfactor2}

Following Ref.~\cite{CCH}, the semileptonic form factors for the $B\to A$
transition are alternatively defined in the following way
\begin{eqnarray} \label{app:FF}
\langle{A}(p, \lambda)|A_\mu|{\overline {B_q}} (p_B)\rangle
 &=& i \frac{2}{m_{B_q} - m_{A}} \varepsilon_{\mu\nu\alpha\beta} \epsilon_{(\lambda)}^{*\nu}
 p_B^\alpha p^{\beta} A^{{B_q} A}(q^2),
\nonumber \\
 \langle A (p,\lambda)|V_\mu|{\overline B_q}(p_B)\rangle
 &=& - \Bigg\{ (m_{B_q} - m_{A}) \epsilon^{(\lambda)*}_{\mu} V_1^{B_q A}(q^2)
 - (\epsilon^{(\lambda)*} \cdot p_B)
(p_B + p)_\mu \frac{V_2^{B_q A}(q^2)}{m_{B_q} - m_{A}}
\nonumber \\
&& - 2 m_{A} \frac{\epsilon^{(\lambda)*}\cdot p_B}{q^2} q^\mu
\left[V_3^{B_q A}(q^2) - V_0^{B_q A}(q^2)\right]\Bigg\}, \nonumber \\
 \end{eqnarray}
where $q = p_B - p$, $V_3^{B_q A}(0) = V_0^{B_q A}(0)$ and
 \begin{eqnarray}
 V_3^{B_q A}(q^2)&=&
 \frac{m_{B_q} - m_{A}}{2 m_{A}} V_1^{B_q A}(q^2) -
 \frac{m_{B_q} + m_{A}}{2 m_{A}} V_2^{B_q A}(q^2)\,,\nonumber\\
 \langle A |\partial_\mu V^\mu | B_q \rangle
 & = & 2i m_A (\epsilon^*p_B) V_0^{B_q A}(q^2).
 \label{app:V30}
 \end{eqnarray}
Following the above parametrization, the numerical results of the light-cone
sum rules for the form factors are listed in Table~\ref{tab:FFinLCSR-2}.
\begin{table}[t]
\caption{Following the parametrization given in Eq.~(\ref{app:FF}), form
factors for $B_{u,d,s} \to  A$ transitions obtained in the light-cone sum rule
calculation are fitted to the 3-parameter form in Eq.~(\ref{eq:FFpara}). Here,
because the decay constants, $f_{^3P_1}$ and $f_{^1P_1}^\perp$, are of the same
sign, the form factors for $B_{(s)}\to 1^1P_1$ and $B_{(s)}\to 1^3P_1$
transitions have opposite signs.} \label{tab:FFinLCSR-2}
\begin{ruledtabular}
\begin{tabular}{| c c c c || c l c c |}
~~~$F$~~~~~
    & $F(0)$~~~~~
    &$a$~~~~~
    & $b$~~~~~~
& ~~~ $F$~~~~~
    & ~~~~~$F(0)$
    & $a$~~~~~
    & $b$~~~~~~
 \\
    \hline
$V_1^{Ba_1}$
    & $0.60\pm0.11$
    & ~~$0.645$
    & $0.250$
&$V_1^{Bb_1}$
    & $-0.32\pm 0.06$
    & ~~$0.748$
    & ~~$0.063$
    \\
$V_2^{Ba_1}$
    & $0.26\pm0.05$
    & $1.48$
    & $1.00~~$
&$V_2^{Bb_1}$
    & $-0.06\pm0.01$
    & ~~$0.539$
    & $1.76$
    \\
$V_0^{Ba_1}$
    & $0.30\pm 0.05$
    & $1.77$
    & $0.926$
&$V_0^{Bb_1}$
    & $-0.39\pm0.07$
    & $1.22$
    & ~~$0.426$
    \\
$A^{Ba_1}$
    & $0.30\pm0.05$
    & $1.64$
    & $0.986$
&$A^{Bb_1}$
    & $-0.16\pm0.03$
    & $1.69$
    & ~~$0.910$
    \\
$V_1^{BK_{1A}}$
    & $0.56\pm0.11$
    & ~~$0.635$
    & $0.211$
&$V_1^{BK_{1B}}$
    & $-0.48^{+0.13}_{-0.08}$
    & ~~$0.729$
    & ~~$0.074$
    \\
$V_2^{BK_{1A}}$
    & $0.25\pm 0.05$
    & $1.51$
    & $1.18~~$
&$V_2^{BK_{1B}}$
    & $-0.10^{+0.03}_{-0.02}$
    & ~~$0.919$
    & ~~$0.855$
    \\
$V_0^{BK_{1A}}$
    & $0.22\pm0.04$
    & $2.40$
    & $1.78~~$
&$V_0^{BK_{1B}}$
    & $-0.45^{+0.12}_{-0.08}$
    & $1.34$
    & ~~$0.690$
    \\
$A^{BK_{1A}}$
    & $0.27\pm0.05$
    & $1.60$
    & $0.974$
&$A^{BK_{1B}}$
    & $-0.22^{+0.06}_{-0.04}$
    & $1.72$
    & ~~$0.912$
    \\
$V_1^{Bf_1}$
    & $0.37\pm0.07$
    & ~~$0.640$
    & $0.153$
&$V_1^{Bh_1}$
    & $-0.21\pm 0.04$
    & ~~$0.612$
    & ~~$0.078$
    \\
$V_2^{Bf_1}$
    & $0.16\pm0.03$
    & $1.47$
    & $0.956$
&$V_2^{Bh_1}$
    & $-0.04\pm0.01$
    & ~~$0.500$
    & $1.63$
    \\
$V_0^{Bf_1}$
    & $0.18\pm 0.03$
    & $1.81$
    & $0.880$
&$V_0^{Bh_1}$
    & $-0.24\pm0.04$
    & $1.16$
    & ~~$0.294$
    \\
$A^{Bf_1}$
    & $0.18\pm0.03$
    & $1.63$
    & $0.900$
&$A^{Bh_1}$
    & $-0.10\pm0.02$
    & $1.54$
    & ~~$0.848$
    \\
$V_1^{Bf_8}$
    & $0.26\pm0.05$
    & ~~$0.644$
    & $0.209$
&$V_1^{Bh_8}$
    & $-0.18\pm 0.03$
    & ~~$0.623$
    & ~~$0.094$
    \\
$V_2^{Bf_8}$
    & $0.11\pm0.02$
    & $1.49$
    & $1.09~~$
&$V_2^{Bh_8}$
    & $-0.03\pm0.01$
    & ~~$0.529$
    & $1.53$
    \\
$V_0^{Bf_8}$
    & $0.12\pm 0.02$
    & $1.84$
    & $0.749$
&$V_0^{Bh_8}$
    & $-0.18\pm0.03$
    & $1.22$
    & ~~$0.609$
    \\
$A^{Bf_8}$
    & $0.13\pm0.02$
    & $1.64$
    & $0.919$
&$A^{Bh_8}$
    & $-0.08\pm0.02$
    & $1.56$
    & ~~$0.827$
    \\
   \hline
$V_1^{B_sK_{1A}}$
    & $0.49\pm0.10$
    & ~~$0.635$
    & $0.211$
&$V_1^{B_sK_{1B}}$
    & $-0.42^{+0.11}_{-0.07}$
    & ~~$0.729$
    & ~~$0.074$
    \\
$V_2^{B_sK_{1A}}$
    & $0.22\pm 0.04$
    & $1.51$
    & $1.18~~$
&$V_2^{B_s K_{1B}}$
    & $-0.09^{+0.03}_{-0.02}$
    & ~~$0.919$
    & ~~$0.855$
    \\
$V_0^{B_s K_{1A}}$
    & $0.19\pm0.04$
    & $2.40$
    & $1.78~~$
&$V_0^{B_s K_{1B}}$
    & $-0.40^{+0.11}_{-0.07}$
    & $1.34$
    & ~~$0.690$
    \\
$A^{B_s K_{1A}}$
    & $0.24\pm0.04$
    & $1.60$
    & $0.974$
&$A^{B_s K_{1B}}$
    & $-0.19^{+0.05}_{-0.03}$
    & $1.72$
    & ~~$0.912$
    \\
$V_1^{B_s f_1}$
    & $0.33\pm0.06$
    & ~~$0.640$
    & $0.153$
&$V_1^{B_s h_1}$
    & $-0.18\pm 0.04$
    & ~~$0.612$
    & ~~$0.078$
    \\
$V_2^{B_s f_1}$
    & $0.14\pm0.03$
    & $1.47$
    & $0.956$
&$V_2^{B_s h_1}$
    & $-0.04\pm0.01$
    & ~~$0.500$
    & $1.63$
    \\
$V_0^{B_s f_1}$
    & $0.16\pm 0.03$
    & $1.81$
    & $0.880$
&$V_0^{B_s h_1}$
    & $-0.21\pm0.04$
    & $1.16$
    & ~~$0.294$
    \\
$A^{B_s f_1}$
    & $0.16\pm0.03$
    & $1.63$
    & $0.900$
&$A^{B_s h_1}$
    & $-0.09\pm0.02$
    & $1.54$
    & ~~$0.848$
    \\
$V_1^{B_s f_8}$
    & $-0.46\pm0.09$~~\,
    & ~~$0.644$
    & $0.209$
&$V_1^{B_s h_8}$
    & ~~$0.32\pm 0.05$
    & ~~$0.623$
    & ~~$0.094$
    \\
$V_2^{B_s f_8}$
    & $-0.19\pm0.03$~~\,
    & $1.49$
    & $1.09~~$
&$V_2^{B_s h_8}$
    & ~~$0.05\pm0.02$
    & ~~$0.529$
    & $1.53$
    \\
$V_0^{B_s f_8}$
    & $-0.21\pm 0.04$~~\,
    & $1.84$
    & $0.749$
&$V_0^{B_s h_8}$
    & ~~$0.32\pm0.05$
    & $1.22$
    & ~~$0.609$
    \\
$A^{B_s f_8}$
    & $-0.23\pm0.04$~~\,
    & $1.64$
    & $0.919$
&$A^{B_s h_8}$
    & ~~$0.14\pm0.03$
    & $1.56$
    & ~~$0.827$
    \\
\end{tabular}
\end{ruledtabular}
\end{table}

\section{Decay amplitudes}\label{app:decay-amplitude}

The differential decay rates for $\bar B_q ^0 \rightarrow A^+ e^- \bar\nu_e$
are given by

\begin{eqnarray}
\frac{d\Gamma(\bar B_q ^0 \rightarrow A^+ e^- \bar\nu_e)} {d E_e\ d q^2}
&=& \frac{G_F^2}{128\pi^3 } |V_{CKM}|^2 \frac{q^2}{m_B^2}\cr
&& \times\bigl [(1-\cos \theta)^2H_-^2+(1+\cos \theta)^2H_+^2+
2(1-\cos^2\theta)H_0^2] ,
\end{eqnarray}
with the helicity amplitudes being
\begin{eqnarray}
H_\pm &=& (m_{B_q}+m_{A})V_1(q^2)\mp
{\tilde\lambda^{1/2}\over m_{B_q}+m_{A}}A(q^2),\cr
 H_0 &=& {1\over2m_{A}(q^2)^{1/2}}\bigl[(m_{B_q}^2-m_{A}^2-q^2)(m_{B_q}+m_{A})V_1(q^2)-
{\tilde\lambda\over m_{B_q}+m_{A}}V_2(q^2)\bigl].
\end{eqnarray}
Here $E_e$ is the electron energy in the $B_q$ rest system. $\theta$ is the
polar angle between the $A$ and $e^-$ in the ($e^-, \bar\nu_e$) system, and is
given by

\begin{eqnarray}
\cos\theta
 =\frac{(m_{B_q}^2-m_A^2+q^2)-4 m_{B_q} E_e}{\tilde\lambda^{1/2}},
\end{eqnarray}
with $\tilde\lambda=(m_{B_q}^2+m^2_{A}-q^2)^2-4m_{B_q}^2 m^2_{A}$.
For a fixed electron energy, $q^2$ varies over the region
$0 \leq q^2\leq q^2_{max}$, where

\begin{eqnarray}
q^2_{max} &=& {2  E_e (m_{B_q}^2-m_A^2 -2m_{B_q} E_e)\over m_{B_q} -2E_e},
\end{eqnarray}
and the related range of $E_e$ is:
$0 \leq E_e \leq (m_{B_q}^2-m_A^2)/(2m_{B_q})$.

\end{document}